\documentclass[journal]{IEEEtran}

\usepackage{amsmath,amssymb,amsfonts,cases,subeqnarray,color}
\usepackage{graphicx}
\usepackage{textcomp}
\usepackage{stfloats}

\def\BibTeX{{\text B\kern-.05em{\sc i\kern-.025em b}\kern-.08em
    T\kern-.1667em\lower.7ex\hbox{E}\kern-.125emX}}
\begin{document}

\title{Physical-Layer Security for Frequency Diverse Array Based Directional Modulation in \\ Fluctuating  Two-Ray Fading Channels}

\author{Qian~Cheng,~\IEEEmembership{Student Member, IEEE},
        Shilian~Wang,~\IEEEmembership{Member, IEEE},
        Vincent~Fusco,~\IEEEmembership{Fellow, IEEE},
        Fanggang~Wang,~\IEEEmembership{Senior Member, IEEE},
        Jiang~Zhu,~\IEEEmembership{Member, IEEE},
        and Chao~Gu

\thanks{The work of Q. Cheng was supported by a scholarship from China Scholarship Council (CSC) under Grant 201803170247. The work of F. Wang was supported in part by the Beijing Natural Haidian Joint Fund under Grant L172020,  in part by the National Natural Science Foundation under Grant 61571034 and Grant U1834210, in part by the Beijing Natural Science Foundation under Grant 4182051, in part by the State Key Laboratory of Rail Traffic Control and Safety under Grant RCS2019ZT011, and in part by the Major Projects of Beijing Municipal Science and Technology Commission under Grant Z181100003218010. \textit {(Corresponding Author: Shilian Wang; Vincent Fusco.)}}
\thanks{Q. Cheng is with the College of Electronic Science, National University of Defense Technology, Changsha 410073, China, and also with the Institute of Electronics, Communications and Information Technology (ECIT), Queen's University Belfast, Belfast BT3 9DT, U.K. (email: chengqian14a@nudt.edu.cn).}
\thanks{S. Wang and J. Zhu are with the College of Electronic Science, National University of Defense Technology, Changsha 410073, China (e-mail: wangsl@nudt.edu.cn, jiangzhu@nudt.edu.cn).}
\thanks{V. Fusco and C. Gu are with the Institute of Electronics, Communications and Information Technology (ECIT), Queen's University Belfast, Belfast BT3 9DT, U.K. (e-mail: v.fusco@ecit.qub.ac.uk,  chao.gu@qub.ac.uk).}
\thanks{F. Wang is with the State Key Laboratory of Rail Traffic Control and Safety, Beijing Jiaotong University, Beijing 100044, China (e-mail: wangfg@bjtu.edu.cn).}

}

\maketitle
\pagestyle{empty}  
\thispagestyle{empty} 

\begin{abstract}
The frequency diverse array (FDA) based directional modulation (DM) technology plays an important role in the implementation of the physical-layer security (PLS) transmission of 5G and beyond communication system. In order to meet the tremendous increase in mobile data traffic, a new design consuming less memory for the FDA-DM-based PLS transmission  is urgently demanded. In this paper, an analytical symmetrical multi-carrier FDA model is proposed in three dimensions, namely, range, azimuth angle, and elevation angle, which differs from the conventional analytical approach with only range and azimuth angle considered. Then, a single-point (SP) artificial noise (AN) aided FDA-DM scheme is  proposed, which  reduces  memory consumption of FDA-DM systems significantly compared with the conventional zero-forcing (ZF) and singular value decomposition (SVD) approaches. Moreover, the PLS performance of the proposed low-memory-consumption FDA-DM scheme is analyzed in fluctuating two-ray (FTR) fading channels for the first time, including bit error rate (BER), secrecy rate (SR), and secrecy outage probability (SOP). More importantly, the closed-form expressions for the lower bound of the average SR and the upper bound of the SOP are derived, respectively. The effectiveness of the analytical expressions is verified by numerical simulations. This work opens a way to lower the memory requirements for the DM-based PLS transmission of 5G and beyond communication system.
\end{abstract}


\section{Introduction}
\IEEEPARstart{P}{hysical-layer} security (PLS) is one of the most important aspects of 5G and beyond wireless communications \cite{Wu_5G_PLS_Review}. Implementing the PLS transmission can result in considerable memory consumptions, which imposes stringent requirements on the 5G nodes or devices. 
In addition, the tremendously increasing mobile data traffic also requires considerable memory consumptions. Therefore, it is highly necessary to lower the memory consumption of the PLS transmission strategy for 5G and beyond communication system.


The directional modulation (DM) technology \cite{Ding_Review_DM}, which is capable of steering the standard baseband symbols along a desired direction while simultaneously distorting the received signals along other directions, has been regarded as a useful PLS transmission strategy for 5G millimeter-wave (mmWave) wireless communications \cite{Wang_DM_mmWave}\cite{Nusenu_FDA_mmWave}. Traditionally,  DM technology is implemented using phased arrays (PA) \cite{Daly_PA_DM}, which only achieves one-dimension security in the direction  while loses security if the eavesdropper is in the same direction  as the legitimate receiver. Compared with the PA-based DM technology, the frequency diverse array (FDA), exhibiting an extra range-dimension dependence apart from angle \cite{Wang_FDA_Review}\cite{Nusenu_FDA_Review}, has been applied into DM implementations to realize two-dimension security in both range and angle.

Specifically, the FDA was first utilized in \cite{Xiong_FDA_DM} to achieve range-angle dependent secure DM transmissions  with fixed linear frequency increments.  The work in \cite{Wang_FDA_DM} utilized the FDA with non-linear  frequency increments to decouple range- and angle-dependent transmit beam patterns for DM transmissions. The FDAs with random and time-modulated frequency increments were exploited for secure DM transmissions in \cite{Hu_Random_FDA_DM} and \cite{Cheng_Time_FDA_DM}, respectively. FDA was also used in \cite{Qiu_AN_FDA_DM}\cite{Lin_FDA_PLS} to establish secure DM transmissions for proximal legitimate user and eavesdropper.  In addition to the single-user FDA-DM schemes  \cite{Xiong_FDA_DM}-\cite{Lin_FDA_PLS}, multi-user FDA-DM schemes were also investigated intensively by means of the spread spectrum technology \cite{Xie_OFDA_MB_DM}, optimization algorithms \cite{Qiu_MB_FDA_DM}, singular value decomposition (SVD) \cite{Cheng_SVD_FDA_DM}, weighted-type  fractional Fourier transform (WFRFT) \cite{Cheng_WFRFT_FDA_DM}, respectively. 

In the DM transmission schemes, artificial noise (AN) plays an important role. Most of the AN-aided DM transmission schemes employ zero-forcing (ZF) method to design the orthogonal precoding matrix to remove the interference of AN for legitimate receivers \cite{Hu_Random_FDA_DM}, \cite{Qiu_AN_FDA_DM}, \cite{Qiu_MB_FDA_DM}, \cite{Hu_AN_MB}-\cite{Xie_AN_MB}. The SVD method provided another way to redesign the orthogonal precoding matrix \cite{Cheng_SVD_FDA_DM}. These ZF or SVD-aided design approaches, however, consume too much memory to store the designed orthogonal matrix and AN. It still remains a challenge to design a secure DM transmission scheme with low memory consumption for 5G and beyond communication system. Moreover, the afore mentioned DM-related works \cite{Xiong_FDA_DM}-\cite{Xie_AN_MB} only considered the line-of-sight (LoS) channels in free space. Regarding the FDA-DM transmission in multipath fading channels, the authors in \cite{Ji_Rayleigh_FDA_DM} and  \cite{Ji_Nakagami_FDA_DM1}\cite{Ji_Nakagami_FDA_DM2} investigated the PLS performance of the FDA-DM communication system in Rayleigh and Nakagami-{\textit m} fading channels, respectively. 

However, on the one hand, the works in \cite{Ji_Rayleigh_FDA_DM}-\cite{Ji_Nakagami_FDA_DM2}   about FDA-DM transmissions in fading channels utilized ZF-based AN method, which demand high memory requirements as well. On the other hand, these conventional fading models like Rayleigh, Rician and Nakagami-{\textit m} fading cannot accurately fit the random small-scale  fluctuations   in real communication environments \cite{Romero_FTR_Conf}. Recently, the fluctuating two-ray (FTR) fading model was proposed in \cite{Romero_FTR_Conf}\cite{Romero_FTR}, which can provide a better fit for small-scale fading measurements in mmWave communications. The authors in \cite{Zhang_FTR}, \cite{Badarneh_Cascaded_FTR} and \cite{Zheng_Squared_FTR} generalized the FTR fading model into arbitrary fading parameter case,  cascaded case, and squared case, respectively. More recently, the secrecy rate (SR), secrecy outage probability (SOP) and symbol error rate (SER) of FTR fading channels were analyzed in \cite{Zeng_FTR_PLS}, \cite{Zhao_SOP_FTR} and \cite{Bilim_SER_PLS} without AN (NoAN), respectively. The power adaption algorithm and wireless-powered UAV relay communication in FTR fading channels were investigated in \cite{Zhao_Power_FTR} and \cite{Zheng_UAV_FTR}, respectively. 

To the best of our knowledge, there is no specific work in the state-of-the-art  that aims to reduce the memory consumption of the FDA-DM scheme for  5G and beyond communications and to analyze the PLS performance of the FDA-DM  scheme  in FTR fading channels. We are the first to make this effort by proposing a low-memory-consumption single-point (SP) AN-aided FDA-DM scheme for 5G and beyond communications, and analyzing its PLS performance in FTR fading channels for the fist time. Overall, the main contributions of our work are  as follows:

\begin{enumerate}
\item Different from the conventional analytical approach which only considers range and azimuth angle dimensions, an analytical model for the symmetrical multi-carrier FDA is proposed in three dimensions, i.e., range, azimuth angle and elevation angle. 

\item Based on the proposed FDA model, a low-memory-consumption FDA-DM scheme is further proposed with the assistance of single-point AN, which significantly outperforms the conventional ZF method \cite{Hu_Random_FDA_DM}, \cite{Qiu_AN_FDA_DM}, \cite{Qiu_MB_FDA_DM}, \cite{Hu_AN_MB}-\cite{Ji_Nakagami_FDA_DM2} and the SVD method \cite{Cheng_SVD_FDA_DM}. The proposed low-memory-consumption FDA-DM scheme provides an efficient strategy to lower the memory requirements for the PLS transmissions of 5G and beyond communications.

\item The bit error rate (BER), SR and SOP performances of the proposed FDA-DM scheme are analyzed in FTR fading channels for the first time. We also derive the closed-form expressions for the lower bound of average SR and the upper bound of SOP. Numerical experiments are conducted to compare the PLS performances of the proposed SP and the conventional ZF \cite{Hu_Random_FDA_DM}, \cite{Qiu_AN_FDA_DM}, \cite{Qiu_MB_FDA_DM}, \cite{Hu_AN_MB}-\cite{Ji_Nakagami_FDA_DM2},  SVD \cite{Cheng_SVD_FDA_DM} and NoAN \cite{Zeng_FTR_PLS}-\cite{Bilim_SER_PLS} methods.

\end{enumerate}

The remainder of this paper is organized as follows. Section II proposes an analytical model of symmetrical multi-carrier FDA in three dimensions. A low-memory-consumption FDA-DM scheme is proposed in Section III with the assistance of single-point AN, where the comparison between the proposed SP method and the conventional ZF and SVD methods is also provided. Section IV analyzes the BER, average SR and  SOP performances of the proposed FDA-DM scheme in FTR fading channels. Numerical results are conducted in Section V in order to verify the advantages of the proposed FDA-DM scheme. Finally, Section VI makes a conclusion for the whole paper and points out the future work.

\emph{Notations}: In this paper, $\imath=\sqrt{-1}$ indicates imaginary unit. The operators ${( \cdot )^{\text{T}}}$ and ${( \cdot )^{\text{H}}}$ represent the transpose and the Hermitian transpose of a matrix. The set of complex numbers is denoted by  $\mathbb{C}$.  The notation $\mathbb{E}(\cdot)$ refers to the expectation of a random variable, while  ${\text {tr}}(\cdot)$ represents the trace of a matrix. In addition,  $\max\{\cdot\}$ and $|\cdot|$ refer to the maximum value of a set of real numbers and the modulus of a complex number, respectively. ${\cal N}(0,{\sigma ^2})$ and ${\cal CN}(0,{\sigma ^2})$ refer to the real and complex Gaussian distributions with zero mean and variance ${\sigma ^2}$, respectively. ${\cal U}[\cdot,\cdot)$ is the uniform distribution. Finally,   the probability function is denoted by   ${\text {Pr}}(\cdot)$.

\section{FDA Model in Three Dimensions}

\begin{figure}
\centering
\includegraphics[angle=0,width=0.45\textwidth]{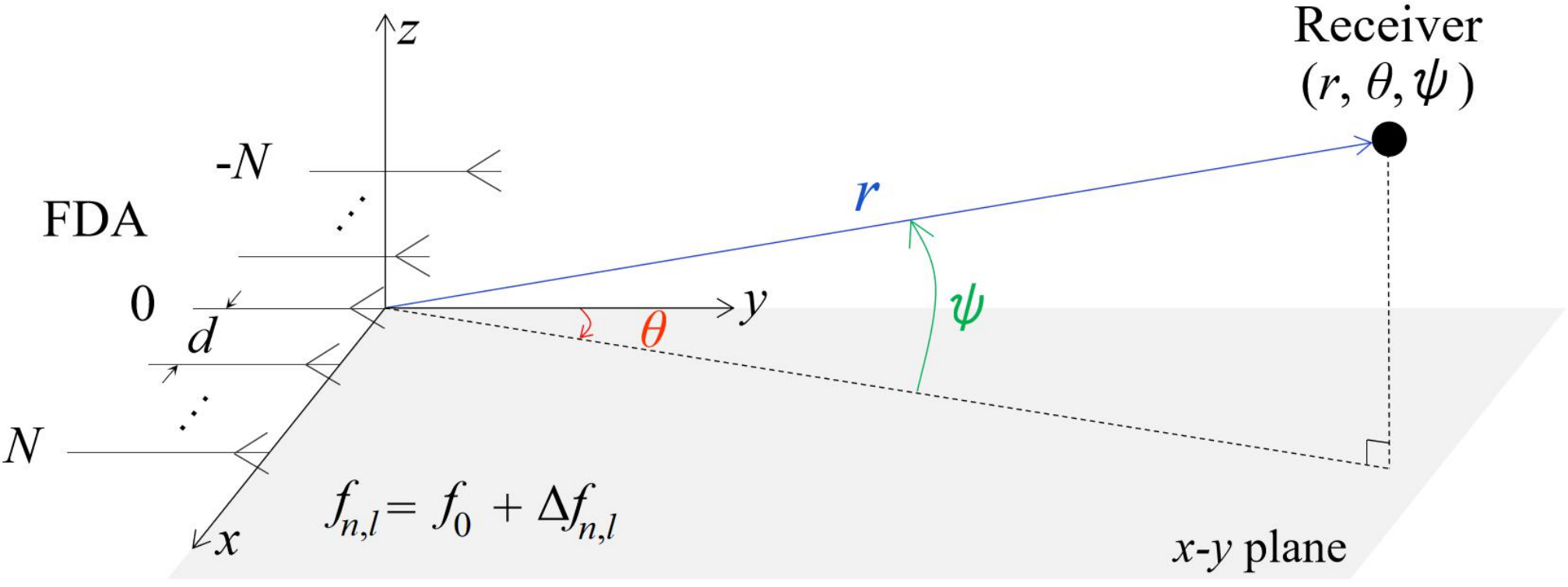}
\caption{The proposed FDA-DM scheme in FTR fading channels.}
\end{figure}

Most of the state-of-the-art  analyzes the FDA only in two dimensions \cite{Wang_FDA_Review}-\cite{Cheng_WFRFT_FDA_DM}, namely, range and azimuth angle. In this paper, we take an extra dimension, elevation angle, into consideration and  establish an analytical FDA model in three dimensions, i.e., range, azimuth angle, and elevation angle.

As shown in Fig. 1, the FDA consists of  $2N+1$ antenna elements  with equal element spacing $d$, which is set as half wavelength of the central carrier. These elements are symmetrically and linearly arrayed on the $x$-axis with the central element located at the coordinate origin. For each element, there are $L$ subcarriers to transmit. The radiated frequency of the $l$-th ($l=0,\cdots,L-1$) subcarrier of the $n$-th ($n=-N,\cdots,0,\cdots,N$) element is designed as 
\begin{equation}
\label{f_n_l}
\begin{aligned}
f_{n,l}&=f_{0}+\Delta f_{n,l}\\
&=f_{0}+\Delta f\ln (|n|+1)\ln(l+1)
\end{aligned}
\end{equation}   
where $f_0$ is the central radiated frequency, $\Delta f$ refers to a fixed frequency increment, and $\Delta f_{n,l}=\Delta f\ln (|n|+1)\ln(l+1)$ represents the frequency increment between the central frequency and the $l$-th subcarrier of the $n$-th element, which satisfies 
\begin{equation}
\label{f_n_constraint}
|\Delta f_{n,l}|\ll f_{0}
\end{equation}

In order to derive the steering vector of the FDA, we consider that each element of the transmitter transmits sinusoidal signals with $L$ subcarriers. The $l$-th subcarrier signal transmitted by the $n$-th element at time $t$ can  be expressed as
\begin{equation}
\label{x_n_l_t}
{x_{n,l}}(t) = e^{\imath2\pi {f_{n,l}}t}
\end{equation}

Let $r$, $\theta$ and $\psi$ represent the range, azimuth angle and elevation angle, respectively, as shown in Fig. 1. For an arbitrary receiver located at $(r,\theta,\psi)$, the overall observed signal in the far field can be written as
\begin{align}
y(r,\theta,\psi )  &= \sum\limits_{n = -N}^{N} \sum\limits_{l = 0}^{L-1}{x_{n,l}}\left( {t - \frac{{{r_n}}}{c}} \right)   \\
\label{y_r_theta_psi1}
& = \sum\limits_{n = -N}^{N} \sum\limits_{l = 0}^{L-1} {\exp \left\{ {\imath{2\pi {f_{n,l}}\left( {t - \frac{{{r_n}}}{c}} \right)}} \right\}}
\end{align}
where $c$ denotes light speed and $r_n$ refers to the path length from the $n$-th element to the observation point. With the far field approximation, $r_n$ can be calculated as
\begin{equation}
\label{far_field}
{r_n} \approx r - nd\sin \theta\cos\psi
\end{equation}
Taking (\ref{f_n_l}) and (\ref{far_field}) into (\ref{y_r_theta_psi1}) yields 
\begin{equation}
\label{y_r_theta_psi2}
\begin{aligned}
&y(r,\theta,\psi ) \approx \\
&  \sum\limits_{n = -N}^{N}  \sum\limits_{l = 0}^{L-1} {\exp \left\{ {\imath { 2\pi \left( {{f_0} + \Delta {f_{n,l}}} \right)\left( {t - \frac{{r - nd\sin \theta \cos\psi}}{c}} \right)} } \right\}} \\
& = \exp \left\{ {\imath2\pi {f_0}\left( {t - \frac{r}{c}} \right)} \right\} \sum\limits_{n = -N}^{N}  \sum\limits_{l = 0}^{L-1}  \exp \left\{ \imath 2\pi \left[\Delta {f_{n,l}}\left( {t - \frac{r}{c}} \right) \right.\right. \\
& ~~~\left. \left. + \frac{1}{c}{ {f_0} nd\sin \theta \cos\psi} + \frac{1}{c}{ \Delta {f_{n,l}}nd\sin \theta\cos\psi }\right]  \right\}
\end{aligned}
\end{equation}   
The constraint in (\ref{f_n_constraint}) implies that the last term in the summation, ${{ \Delta {f_{n,l}}nd\sin \theta\cos\psi }}/{c}$, can be omitted, so (\ref{y_r_theta_psi2}) can be further approximated as
\begin{equation}
\label{y_r_theta_psi3}
\begin{aligned}
& y(r,\theta,\psi ) \approx \exp \left\{ {\imath2\pi {f_0}\left( {t - \frac{r}{c}} \right)} \right\}  \\
& \cdot \sum\limits_{n = -N}^{N} \sum\limits_{l = 0}^{L-1} {\exp \left\{ {\imath 2\pi\left[{ \Delta {f_{n,l}}\left( {t - \frac{r}{c}} \right) + \frac{1}{c}{ {f_0} nd\sin \theta \cos\psi}} \right]} \right\}}
\end{aligned}
\end{equation} 
The terms inside the summation of (\ref{y_r_theta_psi3}) are decided by the geometry and the frequency-offset scheme of the FDA. Therefore, the sub-steering vector caused by the $L$ subcarriers of the $n$-th antenna element can be written as \cite{Shao_MC_FDA} 
\begin{equation}
\label{eq_a_n_r_theta_psi}
\begin{aligned}
{\mathbf{a}}_{n}(r,\theta,\psi)& =\left[ 
e^{\imath 2\pi\left( \Delta f_{n,0}\left (t-\frac{r}{c}\right)+\frac{1}{c}{  f_{0}nd\sin \theta\cos\psi}\right)}  
\ldots \right. \\
&~~~~~e^{\imath 2\pi\left( \Delta f_{n,l}\left (t-\frac{r}{c}\right)+\frac{1}{c}{ f_{0}nd\sin \theta\cos\psi}\right)} 
\ldots \\
&~~~~\left. e^{\imath 2\pi\left( \Delta f_{n,L-1}\left (t-\frac{r}{c}\right)+\frac{1}{c}{ f_{0}nd\sin \theta\cos\psi}\right)} 
\right ]^{\rm{T}}
\end{aligned}
\end{equation}
which is an $L \times 1$ vector. 

Therefore, the overall normalized steering vector of the symmetrical multi-carrier FDA can be calculated as 
\begin{equation}
\label{eq_h_r_theta_psi}
\begin{aligned}
{\mathbf{h}}(r,\theta,\psi)&=\frac {1}{\sqrt{(2N+1)L}} \\
&~~~\cdot\left [{\mathbf{a}}_{-N}^{\rm{T}}(r,\theta,\psi) \cdots {\mathbf{a}}_{n}^{\rm{T}}(r,\theta,\psi) \cdots {\mathbf{a}}_{N}^{\rm{T}}(r,\theta,\psi) \right]^{\rm{T}}
\end{aligned}
\end{equation}

\begin{table*}[!t]
\renewcommand{\arraystretch}{1.5}
\caption{Comparison for Memory Requirements of Different FDA-DM Methods}
\label{table1}
\centering
{
\begin{tabular}{cccc}
\hline
{\bf Items} &{\bf ZF} \cite{Hu_Random_FDA_DM}, \cite{Qiu_AN_FDA_DM}, \cite{Qiu_MB_FDA_DM}, \cite{Hu_AN_MB}-\cite{Ji_Nakagami_FDA_DM2} &{\bf SVD} \cite{Cheng_SVD_FDA_DM} &{\bf Proposed SP}\\ [0.1ex]
\hline
Orthogonal matrix/vector & $\mathbf{P}_2^{\text{ZF}}=\mathbf{I}_{(2N+1)L} - \mathbf{h}_{\text B} \mathbf{h}_{\text B}^{\text{H}}$ & $\mathbf{h}_{\text B}^{\text{H}}=\mathbf{U}\left[\mathbf{D}~\mathbf{0}\right]
\left[\mathbf{V}_{1}~\mathbf{V}_{0}\right]
^{\text{H}}$,~$\mathbf{P}_2^{\text{SVD}}=\mathbf{V}_0$ & $\mathbf{p}_2^{\text {SP}}={\text {Any orthogonal vector of } }\mathbf{h}_{\text B}^{\text{H}}$\\
Size of orthogonal matrix/vector & $(2N+1)L\times (2N+1)L$ & $(2N+1)L\times 2N$ & $(2N+1)L\times 1$\\
Artificial noise & $\mathbf{z}^{\text{ZF}}\in \mathbb{C}^{(2N+1)L\times 1}$ &  $\mathbf{z}^{\text{SVD}}\in \mathbb{C}^{2N\times 1}$ & ${z}^{\text {SP}} \in \mathbb{C}$\\
Size of artificial noise & $(2N+1)L\times 1$ &  $2N\times 1$ & $1$\\
Total size  & $(2N+1)^2L^2+(2N+1)L$ &  $2N(2N+1)L + 2N$ & $(2N+1)L+1$\\
Memory complexity & ${\cal O}(N^2L^2)$ &  ${\cal O}(N^2L)$ & ${\cal O}(NL)$\\
\hline
\end{tabular}
}
\\
\end{table*}

\section {Low-Memory-Consumption FDA-DM Scheme in FTR Fading Channels}

We consider a multi-input single-output single-eavesdropper (MISOSE) wiretap channel model. In this model, a legitimate transmitter (Alice), equipped with a ($2N+1$)-element symmetrical FDA with each element having $L$ subcarriers, intends to deliver confidential information to a legitimate receiver (Bob), while an eavesdropper (Eve) at a different location tries to wiretap the confidential information. As analyzed in Section II, the central element of Alice's FDA is the coordinate origin, meanwhile Bob's and Eve's locations are assumed to be $(r_{\text B},\theta_{\text B},\psi_{\text B})$ and $(r_{\text E},\theta_{\text E},\psi_{\text E})$, respectively. We also assume both Bob's and Eve's channels are in the FTR fading.

\subsection{Alice's Transmit Signal}

The transmit signal of Alice consists of two parts. The first is the normalized baseband modulation symbol $s\in \Omega$ and ${\mathbb E}(|s|^2)=1$, where $\Omega$ is the alphabet of the baseband modulation symbols with the size of $M$. The second is the inserted single-point AN $z$, which follows a complex Gaussian distribution, i.e., $z\sim{\cal CN}(0,1)$.  To match the $(2N+1)L$ FDA subcarriers of Alice, the baseband symbol $s$ and the inserted AN $z$ should be precoded with a normalization vector $\mathbf{p}_1$ and an orthogonal vector $\mathbf{p}_2$, respectively, which require
\begin{equation}
\label{p1_design}
{\mathbf h}^{\text H}_{\text B}{\mathbf p}_1=1
\end{equation}
and
\begin{equation}
\label{p2_design}
{\mathbf h}^{\text H}_{\text B}{\mathbf p}_2=0
\end{equation}
where ${\mathbf h}_{\text B}={\mathbf h}(r_{\text B},\theta_{\text B},\psi_{\text B})$ is the normalized steering vector at Bob's location. 

Therefore, the transmitting signal vector which feeds Alice's $2N+1$ antenna elements can be expressed as
\begin{equation}
\label{x}
{\mathbf x} = \beta_1\sqrt{P_s}{\mathbf p}_1 s +\alpha\beta_2\sqrt{P_s}{\mathbf p}_2 z
\end{equation} 
where ${P_s}$ is the total transmitting power; $\alpha = 1/\sqrt{{\text{tr}}(\mathbf{p}_2\mathbf{p}_2^{\text H})}$ is the power normalization factor for the inserted AN;  $\beta_1$ and $\beta_2$ are power splitting factors for the baseband symbol and the inserted AN, respectively, which satisfy the following constraint,
\begin{equation}
\label{beta1_beta2_constraint}
\beta_{1}^{2}+\beta_{2}^{2}=1
\end{equation} 
For the design of $\mathbf{p}_1$ and $\mathbf{p}_2$, we first review the ZF method \cite{Hu_Random_FDA_DM}, \cite{Qiu_AN_FDA_DM}, \cite{Qiu_MB_FDA_DM}, \cite{Hu_AN_MB}-\cite{Ji_Nakagami_FDA_DM2} and the SVD  method \cite{Cheng_SVD_FDA_DM} briefly, and then propose a low-memory-consumption SP method.

{\textit{1) ZF method:}} The normalization vector is directly designed as the normalized steering vector at Bob's location, i.e., $\mathbf{p}_1^{\text {ZF}} = {\mathbf h}_{\text B}$. In addition, the orthogonal matrix of the ZF method is designed as $\mathbf{P}_2^{\text {ZF}}=\mathbf{I}_{(2N+1)L} - \mathbf{h}_{\text B}\mathbf{h}_{\text B}^{\text{H}} $, which is actually a matrix with the size of $(2N+1)L\times (2N+1)L$. In order to match the orthogonal matrix $\mathbf{P}_2^{\text{ZF}}$, the inserted AN of the ZF method should be randomly valued from $\mathbf{z}^{\text{ZF}} \in \mathbb{C}^{(2N+1)L\times 1}$.

{\textit{2) SVD method:}} Similar to the ZF method, the normalization vector of the SVD method is also designed as $\mathbf{p}_1^{\text {SVD}} = {\mathbf h}_{\text B}$. But for the design of the orthogonal matrix, the SVD method first solves the SVD of $\mathbf{h}_{\text B}^{\text{H}}$, i.e., $\mathbf{h}_{\text B}^{\text{H}}=\mathbf{U}\left[\mathbf{D}~\mathbf{0}\right] \left[\mathbf{V}_{1}~\mathbf{V}_{0}\right] ^{\text{H}}$, from which a null space of $\mathbf{h}_{\text B}^{\text{H}}$ can be obtained. Then, the orthogonal matrix $\mathbf{P}_2^{\text{SVD}}$ can be directly designed as $\mathbf{P}_2^{\text{SVD}}=\mathbf{V}_0$, which is a matrix with the size of $(2N+1)L\times 2N$. Consequently, the inserted AN of the SVD method has to be changed as $\mathbf{z}^{\text{SVD}} \in \mathbb{C}^{2N\times 1}$ in order to match the orthogonal matrix $\mathbf{P}_2^{\text{SVD}}$.

\begin{figure}
\centering
\includegraphics[angle=0,width=0.48\textwidth]{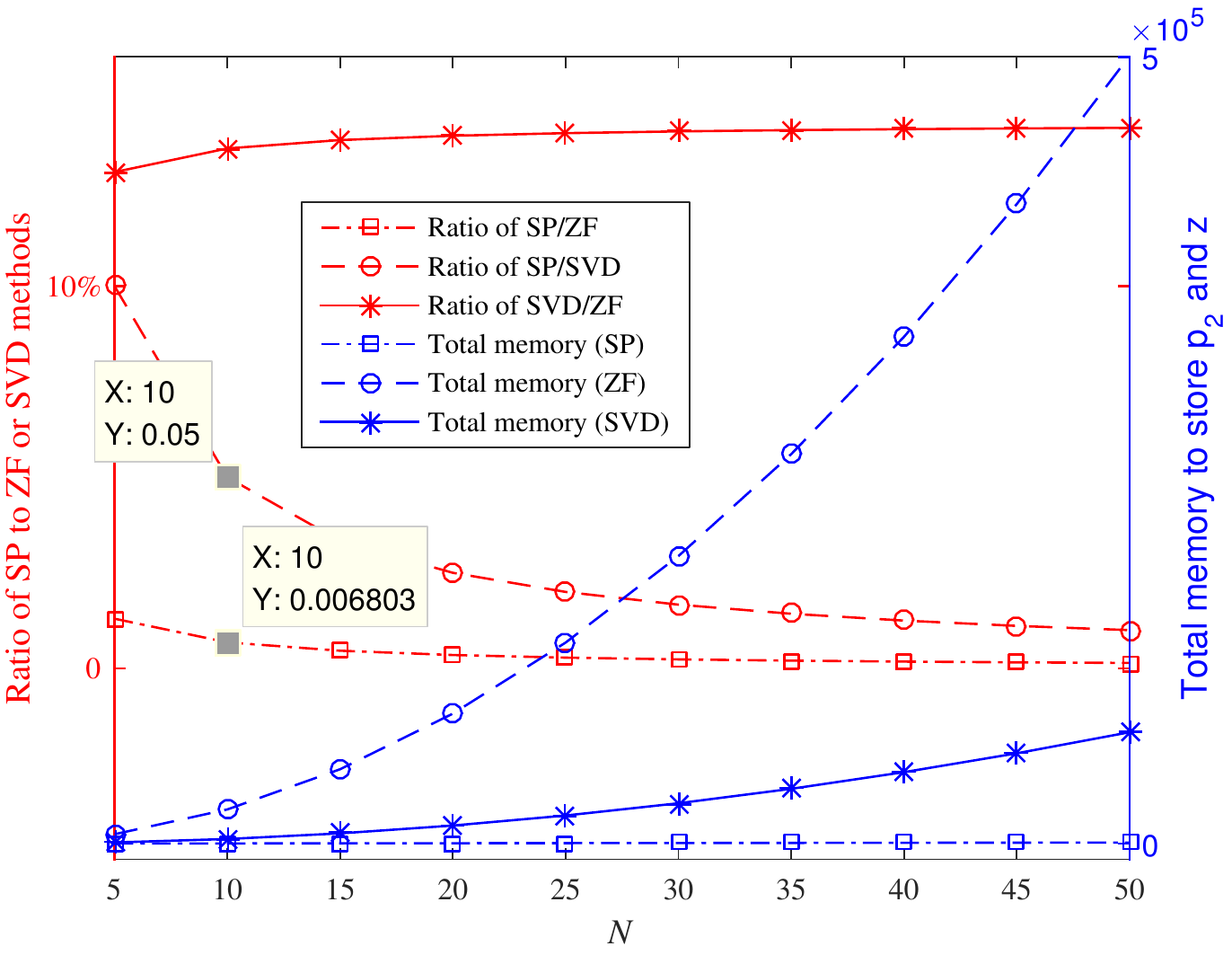}
\caption{Total memory required to store the orthogonal matrix/vector and the AN  versus $N$; and the ratio of the proposed SP method to ZF or SVD methods versus $N$. ($L=7$)}
\end{figure}

{\textit{3) Proposed SP method:}} The normalization vector in this paper is designed as $\mathbf{p}_1^{\text {SP}} = {\mathbf h}_{\text B}$ as well. Different from the ZF and SVD methods which insert an AN vector in the transmit signal, the proposed SP method only requires a single-point AN as shown in (\ref{x}). Therefore, the orthogonal vector ${\mathbf p}_2^{\text {SP}}$ can be designed as an arbitrary orthogonal vector of ${\mathbf h}_{\text B}^{\text{H}}$ rather than a matrix. Comparatively,  ${\mathbf p}_2^{\text {SP}}$ can be an arbitrary column vector of ${\mathbf P}_2^{\text {ZF}}$ or  ${\mathbf P}_2^{\text {SVD}}$. In the following analysis, $\mathbf{p}_2$ specifically refers to the proposed $\mathbf{p}_2^{\text{SP}}$.

{\textit{4) Comparison for the ZF, SVD and SP methods:}} In order to illustrate the advantage of the the proposed SP method, Table I compares these three different design methods in terms of memory consumption to store the orthogonal matrix/vector and the AN. From Table I, the proposed SP method reduces the memory complexity from ${\cal O}(N^2L^2)$ and ${\cal O}(N^2L)$ to ${\cal O}(NL)$, which significantly outperforms the ZF and SVD methods. 

In addition, Fig. 2 and Fig. 3 depict the numerical results for the total memory required and the ratio of SP method to ZF or SVD methods versus $N$ and $L$, respectively, which verify the excellent advantage of low memory consumption for the proposed SP method. For example, when $N=10$ and $L=7$, the proposed SP method only requires approximately $0.68\%$ total memory of the ZF method or $5\%$ total memory of the SVD method.  Fig. 4 and Fig. 5 further show the total memory consumptions and the corresponding ratios versus $N$ and $L$, respectively, from which it demonstrates that the proposed SP method can save much more memory than the conventional ZF and SVD methods as well.

\begin{figure}
\centering
\includegraphics[angle=0,width=0.48\textwidth]{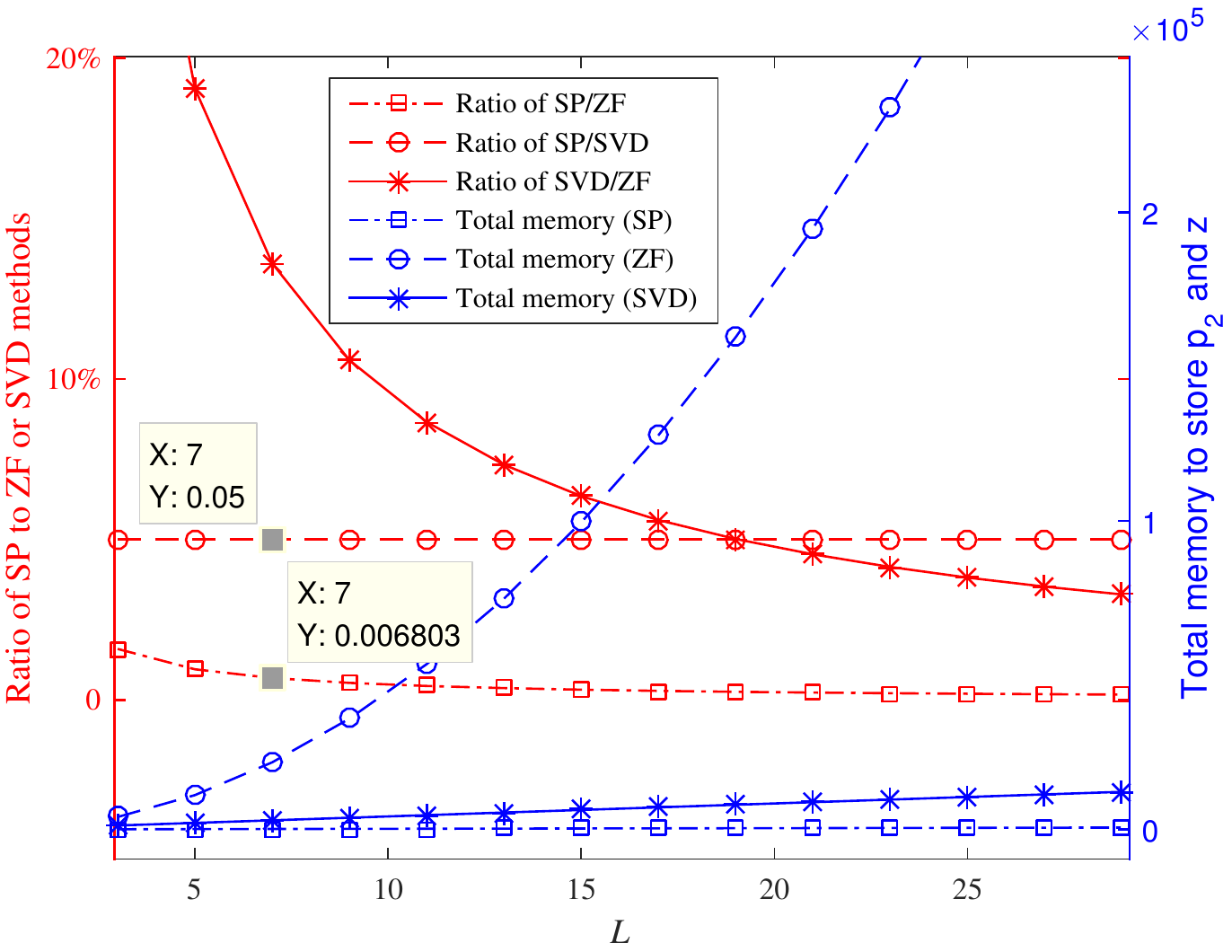}
\caption{Total memory required to store the orthogonal matrix/vector and  the AN  versus $L$; and the ratio of the proposed SP method to ZF or SVD methods versus $L$. ($N=10$)}
\end{figure}

\subsection{Bob's and Eve's Signals}

After Alice transmits the signal, the received signal of Bob located at $(r_{\text B},\theta_{\text B},\psi_{\text B})$ can be written as 
\begin{align}
\label{y_B1}
y(r_{\text B},\theta_{\text B},\psi_{\text B}) 
&= \epsilon_{\text B}{\mathbf h}_{\text B}^{\text H} {\mathbf x}  + {\xi_{\text B}}\\
\label{y_B2}
& = \epsilon_{\text B} \beta_1\sqrt{P_s}{\mathbf h}_{\text B}^{\text H}{\mathbf p}_1 s + \epsilon_{\text B}\alpha\beta_2\sqrt{P_s}{\mathbf h}_{\text B}^{\text H}{\mathbf p}_2 z + {\xi_{\text B}}\\
\label{y_B3}
& =  \epsilon_{\text B}\beta_1\sqrt{P_s} s + {\xi_{\text B}}
\end{align}
where ${\xi_{\text B}} $ is the complex  additive white Gaussian noise (AWGN) with zero mean and variance $\delta_{\text B}^2$, i.e., ${\xi_{\text B}} \sim {\cal{CN}}(0, \delta_{\text B}^2)$. In addition, $\epsilon_{\text B}$ represents the FTR fading coefficient which is defined as \cite{Zhang_FTR}
\begin{equation}
\label{epsilon_B}
\epsilon_{\text B} = \sqrt{\zeta_{\text B}}U_{\text B}e^{\imath\varphi_{\text B}}+\sqrt{\zeta_{\text B}}V_{\text B}e^{\imath\vartheta_{\text B}}+X_{\text B}+\imath Y_{\text B}
\end{equation} 
where ${\zeta_{\text B}}$ is a Gamma distributed random variable with zero mean and probability density function given by
\begin{equation}
\label{pdf_zeta}
f_{\zeta_{\text B}}(\zeta_{\text B})=\frac{m_{\text B}^{m_{\text B}} \zeta_{\text B}^{m_{\text B}-1}}{\Gamma(m_{\text B})}e^{-m_{\text B}\zeta_{\text B}}
\end{equation}
Moreover, $U_{\text B}$ and $V_{\text B}$ are constant amplitudes with specular components modulated by a Nakagami-$m_{\text B}$ random variable. $\varphi_{\text B}$ and $\vartheta_{\text B}$ are statistically independent and uniformly distributed random phases, i.e., $\varphi_{\text B},\vartheta_{\text B}\sim {\cal U}[0,2\pi)$. $X_{\text B} + \imath Y_{\text B}$ refers to the diffuse component with $X_{\text B}$  and $Y_{\text B}$ following a Gaussian distribution, i.e., $X_{\text B},Y_{\text B}\sim {\cal N} (0,\sigma_{\text B}^2) $. The FTR fading parameters can be calculated by $K_{\text B} = \frac{U_{\text B}^2 + V_{\text B}^2}{2\sigma_{\text B}^2}$ and $\Delta_{\text B}=\frac{2U_{\text B}V_{\text B}}{U_{\text B}^2 + V_{\text B}^2}$. It can be observed from (\ref{y_B3}) that only the useful signal is left for Bob while the inserted AN has been removed, which guarantees the effective transmission between Alice and Bob.

\begin{figure}
\centering
\includegraphics[angle=0,width=0.48\textwidth]{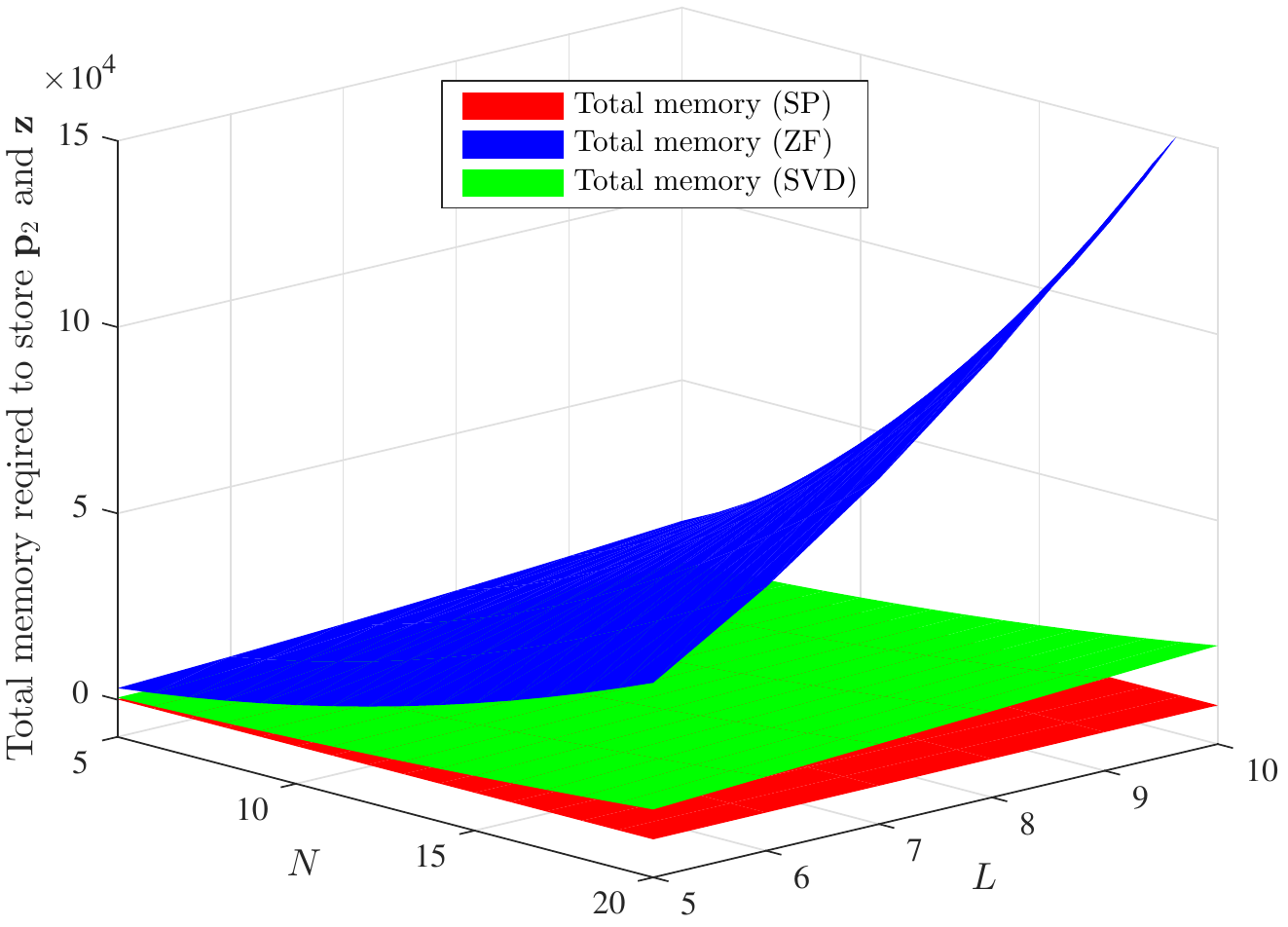}
\caption{Total memory required to store the orthogonal matrix/vector and the AN  of the proposed SP method and the conventional ZF and SVD methods versus $N$ and $L$.}
\end{figure}

Similarly, the received signal of Eve located at $(r_{\text E},\theta_{\text E},\psi_{\text E})$ can be expressed as
\begin{align}
\label{y_E1}
y(r_{\text E},\theta_{\text E},\psi_{\text E}) 
&=  \epsilon_{\text E}{\mathbf h}_{\text E}^{\text H} {\mathbf x} + \xi_{\text E}\\
\label{y_E2}
& = {\epsilon_{\text E}\beta_1\sqrt{P_s}{\mathbf h}_{\text E}^{\text H}{\mathbf p}_1 s}
 + {\epsilon_{\text E}\alpha\beta_2\sqrt{P_s}{\mathbf h}_{\text E}^{\text H}{\mathbf p}_2 z} + \xi_{\text E}\\
 \label{y_E3}
& = \underbrace{\epsilon_{\text E}\beta_1\sqrt{P_s}\rho_1 s}_{\text {Distorted Signal}} + \underbrace{\epsilon_{\text E}\alpha\beta_2\sqrt{P_s}\rho_2 z}_{\text {Artificial Noise}} + \underbrace{ \xi_{\text E}}_{\text {AWGN}}
\end{align}
where ${\xi_{\text E}} \sim {\cal{CN}}(0, \delta_{\text E}^2)$ indicates the complex  AWGN, ${\mathbf h}_{\text E} = {\mathbf h}( r_{\text E},\theta_{\text E},\psi_{\text E})$ refers to Eve's normalized steering vector, $\rho_1 = {\mathbf h}_{\text E}^{\text H}{\mathbf p}_1$, and $\rho_2 = {\mathbf h}_{\text E}^{\text H}{\mathbf p}_2$. Additionally, the fading coefficient $\epsilon_{\text E}$ is defined as 
\begin{equation}
\label{epsilon_B}
\epsilon_{\text E} = \sqrt{\zeta_{\text E}}U_{\text E}e^{\imath\varphi_{\text E}}+\sqrt{\zeta_{\text E}}V_{\text E}e^{\imath\vartheta_{\text E}}+X_{\text E}+\imath Y_{\text E}
\end{equation} 
which undergoes the FTR fading with the parameters $(m_{\text E},K_{\text E},\Delta_{\text E},\sigma_{\text E}^2)$. 

It is worth noting that Eve's received signal in (\ref{y_E3}) consists of three items. The first is the useful signal distorted by $\rho_1$ and the second is the inserted AN. Both can be regarded as interference for Eve, thereby guaranteeing the PLS transmission between Alice and Bob.

\begin{figure}
\centering
\includegraphics[angle=0,width=0.48\textwidth]{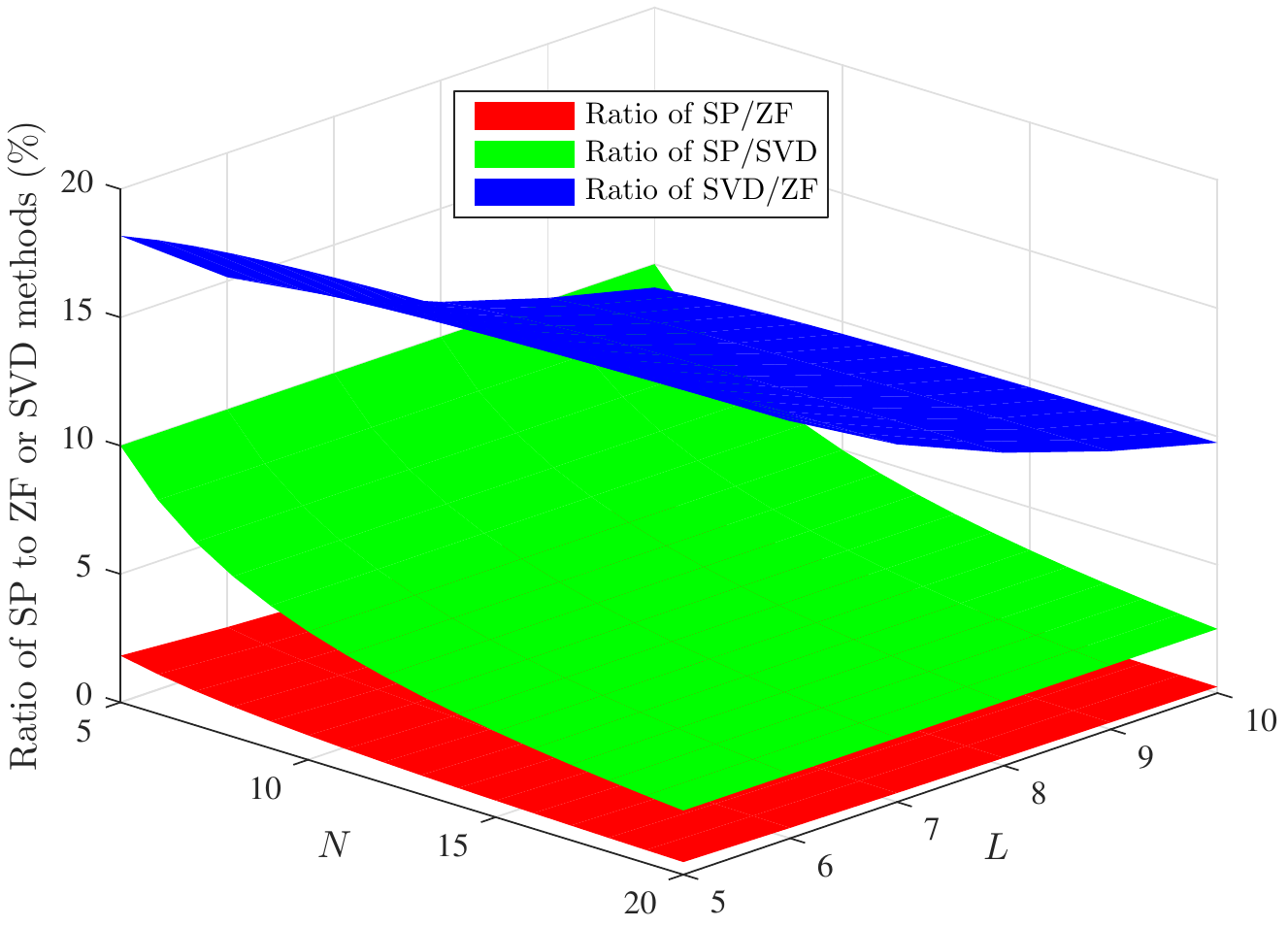}
\caption{The ratio of total memory of the SP method  to that of the ZF or SVD methods  versus $N$ and $L$.}
\end{figure}

\section{Physical-Layer Security Analysis}

In this section, we will analyze the PLS performances of the proposed FDA-DM model including BER, SP and SOP, which are important metrics to measure the performances of DM transmission systems \cite{Ding_Metrics_DM}.

\subsection{Bit Error Rate}

To acquire the BER formula, we consider the case of no fading, which means $\epsilon_{\text B}=\epsilon_{\text E}=1$. As a matter of fact, when the fading is considered, the derived BER formula can refer to \cite{Bilim_SER_PLS} as long as the channel state information (CSI) \cite{Yomo_CSI_Estimation}\cite{Wilkinson_CSI_Estimation} is fully estimated at Bob.

According to (\ref{y_B3}), the received signal of Bob is simply the summation of the useful signal and AWGN, and the average signal-to-noise ratio (SNR) of Bob can be written as
\begin{equation}
\label{SNR_B}
\gamma_{\text B} = \frac{\beta_1^2 P_s {\mathbb E}(|s|^2)}{\delta_{\text B}^2 }
\end{equation} 
For an $M$-ary baseband modulation, the SNR per bit can be calculated by
\begin{equation}
\label{SNR_bit}
\gamma_{\text {bit}} = \frac{\beta_1^2 P_s {\mathbb E}(|s|^2)}{\delta_{\text B}^2 \log_2 M} = \frac{\gamma_{\text B}}{\log_2 M}
\end{equation} 
Therefore, if PSK modulation is adapted, the BER formula for the proposed FDA-DM system can be calculated by \cite{Proakis_Digital_Commu}
\begin{equation}
\label{BER}
\begin{aligned}
P_e & \approx \frac{2}{\log_2 M}Q\left(\sqrt{2\gamma_{\text {bit}} \log_2 M}\sin\frac{\pi}{M}\right) \\
& = \frac{2}{\log_2 M}Q\left(\sqrt{2\gamma_{\text B}}\sin\frac{\pi}{M}\right)
\end{aligned}
\end{equation} 
where $Q(u) = {1 \mathord{\left/
 {\vphantom {1 {\sqrt {2\pi } }}} \right.
 \kern-\nulldelimiterspace} {\sqrt {2\pi } }}\int_u^\infty  {\exp \{  - {{{u^2}} \mathord{\left/
 {\vphantom {{{u^2}} 2}} \right.
 \kern-\nulldelimiterspace} 2}\} } du$ is the tail distribution function of the standard normal distribution.

\subsection{Secrecy Rate}

When the FTR fading is considered, we can rewrite the SNR of Bob as
\begin{equation}
\label{gamma_B}
\gamma_{\text B} = \frac{|\epsilon_{\text B}|^2\beta_1^2P_s{\mathbb E}(|s|^2)}{\delta_{{\text B}}^2}  = \beta_1^2 \lambda_{\text B}
\end{equation} 
where $\lambda_{\text B} = {|\epsilon_{\text B}|^2P_s{\mathbb E}(|s|^2)}/{\delta_{{\text B}}^2}$ is Bob's SNR of the FTR fading channel without splitting AN, the average of which can be calculated by ${\bar\lambda}_{\text B} = {{\mathbb E}(|\epsilon_{\text B}|^2)P_s{\mathbb E}(|s|^2)}/{\delta_{{\text B}}^2} = 2\sigma_{\text B}^2(1+K_{\text B}){P_s}/{\delta_{{\text B}}^2}$ \cite{Zhang_FTR}.

According to (\ref{y_E3}), we can write the signal to inference-plus-noise (SINR) of Eve as 
\begin{equation}
\label{gamma_E1}
\begin{aligned}
\gamma_{\text E} 
& = \frac{|\epsilon_{\text E}|^2\beta_1^2|\rho_1|^2P_s{\mathbb E}(|s|^2)/\delta_{\text E}^2}{|\epsilon_{\text E}|^2\alpha^2\beta_2^2|\rho_2|^2P_s{\mathbb E}(|z|^2)/\delta_{\text E}^2 + 1} 
\end{aligned}
\end{equation}
Let $\lambda_{\text E} = {|\epsilon_{\text E}|^2P_s{\mathbb E}(|s|^2)}/{\delta_{{\text E}}^2}$, which indicates Eve's SNR of FTR fading channels without splitting AN, the average of which can be calculated by ${\bar\lambda}_{\text E} = {{\mathbb E}(|\epsilon_{\text E}|^2)P_s{\mathbb E}(|s|^2)}/{\delta_{{\text E}}^2} = 2\sigma_{\text E}^2(1+K_{\text E}){P_s}/{\delta_{{\text E}}^2}$. Moreover, taking the assumption ${\mathbb E}(|s|^2)={\mathbb E}(|z|^2)=1$ into (\ref{gamma_E1}) yields
\begin{equation}
\label{gamma_E2}
\gamma_{\text E} = \frac{\beta_1^2|\rho_1|^2 \lambda_{\text E}}{\alpha^2\beta_2^2|\rho_2|^2 \lambda_{\text E} + 1} = \frac{\eta\lambda_{\text E}}{\mu\lambda_{\text E} + 1}
\end{equation}
where $\eta = \beta_1^2|\rho_1|^2$ and $\mu = \alpha^2\beta_2^2|\rho_2|^2$.

\begin{figure*}[htp]
\centering
\includegraphics[angle=0,width=0.99\textwidth]{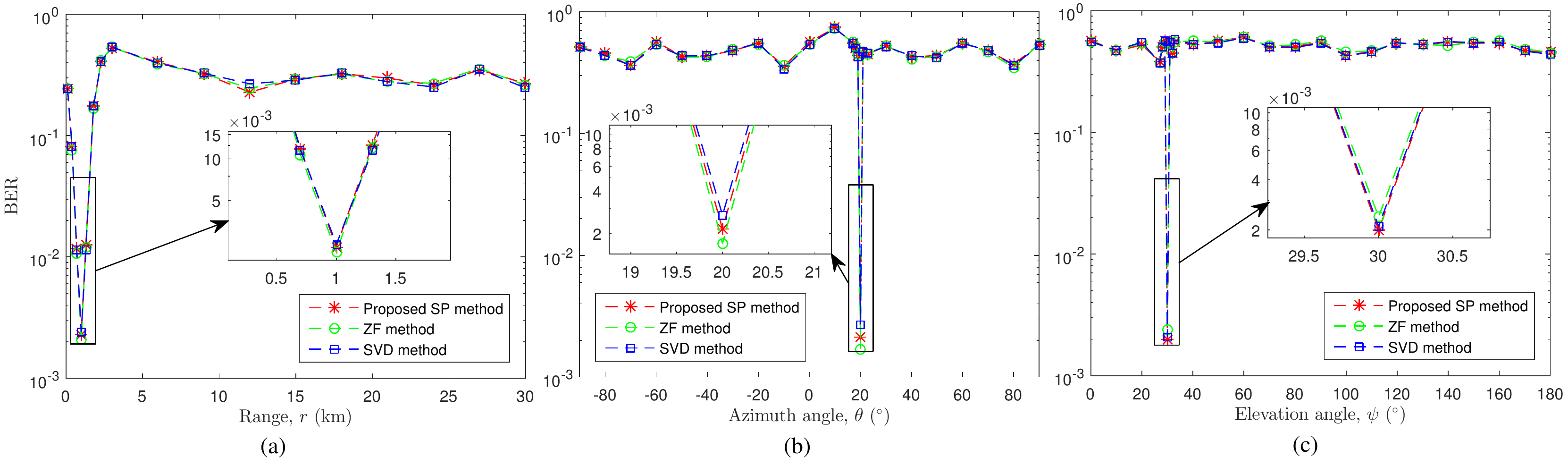}
\caption{BER versus (a) range $r$ with $\theta=\theta_{\text B}=20^\circ$ and $\psi=\psi_{\text B}=30^\circ$, (b) azimuth angle $\theta$ with $r=r_{\text B}=1$ km and $\psi=\psi_{\text B}=30^\circ$, and (c) elevation angle $\psi$ with $r=r_{\text B}=1$ km and $\theta=\theta_{\text B}=20^\circ$. ($f_0=30$ GHz, $\Delta f=20$ kHz, $N=10$, $L=7$, $\beta_1 = 0.9$, QPSK, SNR $=10$ dB)}
\end{figure*}

According to \cite{Zhang_FTR}, the probability density function (PDF) and cumulative distribution function (CDF) of $\lambda_i$ ($i\in\{{\text B}, {\text E}\}$) can be written as
\begin{equation}
\label{pdf_lambda_i}
 f_{\lambda_{i}}(x) = \frac{m_{i}^{m_{i}}}{\Gamma(m_{i})} \sum\limits_{j_{i}=0}^{\infty}\frac{K_{i}^{j_{i}}d_{j_{i}} }{{j_{i}}!{j_{i}}!}\frac{x^{j_{i}}}{(2\sigma_{i}^2)^{{j_{i}} +1}} \exp\left(-\frac{x}{2\sigma_{i}^2}\right)
\end{equation} 
and
\begin{equation}
\label{cdf_lambda_i}
F_{\lambda_{i}}(x) = \frac{m_{i}^{m_{i}}}{\Gamma(m_{i})} \sum\limits_{j_{i}=0}^{\infty} \frac{K_{i}^{j_{i}}d_{j_{i}}}{{j_{i}}!{j_{i}}!}\Upsilon \left({j_{i}}+1,\frac{x}{2\sigma_{i}^2}\right)
\end{equation} 
where ${j_{i}}!$ denotes the factorial of the integer ${j_{i}}$; $\Gamma(\cdot)$ and $\Upsilon(\cdot,\cdot)$ refer to the ordinary Gamma function [39, Eq. (8.310.1)] and  the lower  incomplete Gamma function [39, Eq. (8.350.1)], respectively. In addition, the term $d_{j_{i}}$ in (\ref{pdf_lambda_i}) and (\ref{cdf_lambda_i}) is expressed as
\begin{equation}
\label{d_j_i}
\begin{aligned}
d_{j_{i}} & \triangleq \sum\limits_{k=0}^{j_{i}}{{j_{i}} \choose k}\left(\frac{\Delta_{i}}{2}\right)^{k}\sum\limits_{l=0}^{k}{k \choose l}\Gamma(j_{i}+m_{i}+2l-k)\\
& ~~~\cdot e^{\imath\frac{(2l-k)\pi }{2}}\bigg[(m_{i}+K_{i})^2 - (K_{i}\Delta_{i})^2\bigg]^{-\frac{j_{i}+m_{i}}{2}}\\
& ~~~\cdot {\cal L}_{j_{i}+m_{i} -1}^{k-2l}\left(\frac{m_{i}+K_{i}}{\sqrt{(m_{i}+K_{i})^2 - (K_{i}\Delta_{i})^2}}\right)
\end{aligned}
\end{equation}
where ${k \choose l}$ denotes $k$ chooses $l$, and ${\cal L}_{\cdot}^{\cdot}(\cdot)$ denotes the associated Legendre function of the first kind [39, Eq. (8.702)]. By observing (\ref{gamma_B}), the PDF and CDF of $\gamma_{\text B}$ can be easily calculated by (\ref{pdf_gamma_B}) and (\ref{cdf_gamma_B}), respectively
\begin{equation}
\label{pdf_gamma_B}
f_{\gamma_{\text {B}}}(x) = \frac{1}{\beta_1^2}f_{\lambda_{\text B}}\left(\frac{x}{\beta_1^2}\right)
\end{equation} 
\begin{equation}
\label{cdf_gamma_B}
F_{\gamma_{\text {B}}}(x) = F_{\lambda_{\text B}}\left(\frac{x}{\beta_1^2}\right)
\end{equation} 

Before deriving the PDF and CDF of $\gamma_{\text E}$, we first point out a fact that the lower and upper bounds of $\gamma_{\text E}$ are $0$ and $ \tau= \eta/\mu$, respectively, which can be directly obtained by replacing $\lambda_{\text E}\rightarrow 0$ and $\lambda_{\text E} \rightarrow \infty$ into (\ref{gamma_E2}). Therefore, the CDF of $\gamma_{\text E}$ can be acquired by
\begin{align}
\label{cdf_gamma_E1}
F_{\gamma_{\text E}}(x) 
&={ \text{Pr}}({\gamma_{\text E}}\leqslant x) \\
\label{cdf_gamma_E2}
& ={ \text{Pr}}\left(\frac{\eta{\lambda_{\text E}}}{\mu{\lambda_{\text E}} + 1}\leqslant x\right) \\
\label{cdf_gamma_E3}
& ={ \text{Pr}}\left({\lambda_{\text E}} \leqslant \frac{x}{\eta-\mu x}\right)  \\
\label{cdf_gamma_E4}
& = 
\left\{
\begin{matrix}
 F_{\lambda_{\text E}}\left( \frac{x}{\eta-\mu x}\right), & 0< x < \tau\\
1, & x\geqslant\tau
\end{matrix}
\right.
\end{align}
Consequently, the PDF of  $\gamma_{\text E}$ can be calculated by
\begin{align}
\label{pdf_gamma_E1}
f_{\gamma_{\text E}}(x) &=  \frac{dF_{\gamma_{\text E}}(x)}{dx} \\
\label{pdf_gamma_E2}
& = 
\left\{
\begin{matrix}
 \frac{\eta}{(\eta-\mu x)^2} f_{\lambda_{\text E}}\left( \frac{x}{{\eta} - \mu x}\right) , & 0< x < \tau\\
0, & x\geqslant\tau
\end{matrix}
\right.
\end{align}
Using the PDFs and CDFs of Bob's and Eve's SINRs,  the instantaneous secrecy rate can be defined as
\begin{equation}
\label{R_s}
R_s(\gamma_{\text B},\gamma_{\text E}) = \Big[\log _2 (1+\gamma_{\text B}) - \log _2(1+\gamma_{\text E})\Big]^{+}
\end{equation} 
where $[\cdot]^{+} = \max\{\cdot,0\}$. If we further assume Bob's and Eve's channels experience independent fading, the average secrecy rate can be calculated by \cite{Zeng_FTR_PLS}
\begin{equation}
\label{average_R_s}
\begin{aligned}
\bar{R}_s(\gamma_{\text B},\gamma_{\text E}) 
& = \int\nolimits_{0}^{\infty}\int\nolimits_{0}^{\infty}R_s(\gamma_{\text B},\gamma_{\text E})f(\gamma_{\text B},\gamma_{\text E})d\gamma_{\text B}d\gamma_{\text E}\\
& = \underbrace{\frac{1}{\ln 2}\int\nolimits_{0}^{\infty}\ln(1+\gamma_{\text B})f_{\gamma_{\text B}}(\gamma_{\text B})F_{\gamma_{\text E}}(\gamma_{\text B})d\gamma_{\text B} }_{{\cal I}_1} \\
&~~~ +  \underbrace{\frac{1}{\ln 2}\int\nolimits_{0}^{\infty}\ln(1+\gamma_{\text E})f_{\gamma_{\text E}}(\gamma_{\text E})F_{\gamma_{\text B}}(\gamma_{\text E})d\gamma_{\text E} }_{{\cal I}_2}\\
&~~~ - \underbrace{\frac{1}{\ln 2}\int\nolimits_{0}^{\infty}\ln(1+\gamma_{\text E})f_{\gamma_{\text E}}(\gamma_{\text E})d\gamma_{\text E} }_{{\cal I}_3}
\end{aligned}
\end{equation}
where $f(\gamma_{\text B},\gamma_{\text E})=f_{\gamma_{\text B}}(\gamma_{\text B})f_{\gamma_{\text E}}(\gamma_{\text E})$ is  the joint PDF of ${\gamma_{\text B}}$ and ${\gamma_{\text E}}$.

Before deriving the secrecy rate, we define
\begin{equation}
\label{Psi_v_tau}
\begin{aligned}
& \Psi(v_1,v_2,v_3,v_4,v_5,\tau) \\
& = \int\nolimits_{0}^{\tau}\ln ^{v_1}(1+t)\frac{t^{v_2}}{(\tau - t)^{v_3}}\exp\left\{-{v_4}t-\frac{v_5 t}{\tau - t}\right\}dt
\end{aligned}
\end{equation}
of which a special case is [29, Eq. (10)]
\begin{align} 
\label{S_u_v1}
{ \cal S}(u,v) & =  \Psi(1,u-1,0,v,0,\infty) \\
\label{S_u_v2}
& = \int\nolimits_{0}^{\infty}\ln (1+t){t^{u-1}}\exp\left\{-{v}t\right\}dt \\
\label{S_u_v3}
& = (u-1)!e^{v}\sum\limits_{k=1}^{u}\frac{\Gamma(-u+k,v)}{v^{k}}
\end{align} 
Then, the average secrecy rate can be obtained  in Lemma 1 by substituting (\ref{pdf_gamma_B}), (\ref{cdf_gamma_B}), (\ref{cdf_gamma_E4}) and (\ref{pdf_gamma_E2}) into (\ref{average_R_s}).

{\textit {Lemma 1:}} The average secrecy rate of the proposed FDA-DM system in FTR fading is given by
\begin{equation}
\label{average_R_s_lemma}
\bar{R}_s(\gamma_{\text B},\gamma_{\text E}) 
 = {{\cal I}_1} + {{\cal I}_2} - {{\cal I}_3}
\end{equation}

\newcounter{cnt1}
\setcounter{cnt1}{\value{equation}}
\setcounter{equation}{47}
\begin{figure*}[hbp]
\hrulefill
\begin{equation}
\label{I_1_lemma}
\begin{aligned}
{{\cal I}_1} 
& =  \frac{m_{\text B}^{m_{\text B}}{m_{\text E}^{m_{\text E}}}}{{\ln 2}\Gamma(m_{\text B}){\Gamma(m_{\text E})}} 
\sum\limits_{j_{\text B}=0}^{\infty}\sum\limits_{j_{\text E}=0}^{\infty}
\frac{K_{\text B}^{j_{\text B}}d_{j_{\text B}} {K_{\text E}^{j_{\text E}}d_{j_{\text E}}}}{{j_{\text B}}!{j_{\text B}}!{{j_{\text E}}!} {(2\beta_1^2\sigma_{\text B}^2)^{{j_{\text B}} +1}} }  \left(\chi  
-  \sum\limits_{n=0}^{j_{\text E}} \frac{1}{n!({2\mu\sigma_{\text E}^2})^n}
    \Psi(1,j_{\text B}+n,n,\frac{1}{2\beta_1^2\sigma_{\text B}^2},\frac{1}{2\mu\sigma_{\text E}^2},\tau)\right )\\
&~~~+ \frac{m_{\text B}^{m_{\text B}}}{{\ln 2}\Gamma(m_{\text B})} 
\sum\limits_{j_{\text B}=0}^{\infty}\frac{K_{\text B}^{j_{\text B}}d_{j_{\text B}} }{{j_{\text B}}!{j_{\text B}}!{(2\beta_1^2\sigma_{\text B}^2)^{{j_{\text B}} +1}}}
\left({ \cal S}(j_{\text B}+1,\frac{1}{2\beta_1^2\sigma_{\text B}^2})  
 - \chi\right)
\end{aligned}
\end{equation}
\end{figure*}
\setcounter{equation}{\value{cnt1}}
\setcounter{equation}{48}

\newcounter{cnt2}
\setcounter{cnt2}{\value{equation}}
\setcounter{equation}{48}
\begin{figure*}[hbp]
\begin{equation}
\label{I_2_lemma}
\begin{aligned}
{{\cal I}_2} 
& = \frac{m_{\text B}^{m_{\text B}}{m_{\text E}^{m_{\text E}}}}{{\ln 2}\Gamma(m_{\text B}){\Gamma(m_{\text E})}}
\sum\limits_{j_{\text B}=0}^{\infty} \sum\limits_{j_{\text E}=0}^{\infty} 
\frac{K_{\text B}^{j_{\text B}}d_{j_{\text B}}{K_{\text E}^{j_{\text E}}d_{j_{\text E}} } \tau}{{j_{\text B}}!{{j_{\text E}}!{j_{\text E}}!} (2\mu\sigma_{\text E}^2)^{{j_{\text E}} +1} }\\
& ~~~\cdot \left(  
 \Psi(1,j_{\text E},j_{\text E}+2,0,\frac{1}{2\mu\sigma_{\text E}^2},\tau) -   \sum\limits_{n=0}^{j_{\text B}} \frac{1}{n!(2\beta_1^2\sigma_{\text B}^2)^n}
  \Psi(1,j_{\text E}+n,j_{\text E}+2,\frac{1}{2\beta_1^2\sigma_{\text B}^2},\frac{1}{2\mu\sigma_{\text E}^2},\tau)\right)
\end{aligned}
\end{equation}
\end{figure*}
\setcounter{equation}{\value{cnt2}}
\setcounter{equation}{49}

\newcounter{cnt3}
\setcounter{cnt3}{\value{equation}}
\setcounter{equation}{49}
\begin{figure*}[hbp]
\hrulefill
\begin{equation}
\label{I_3_lemma}
{{\cal I}_3} 
 = \frac{m_{\text E}^{m_{\text E}}}{{\ln 2}\Gamma(m_{\text E})} \sum\limits_{j_{\text E}=0}^{\infty}
\frac{K_{\text E}^{j_{\text E}}d_{j_{\text E}} \tau}{{j_{\text E}}!{j_{\text E}}!(2\mu\sigma_{\text E}^2)^{{j_{\text E}} +1}}
\Psi(1,{j_{\text E}},{j_{\text E}+2},0,\frac{1}{2\mu\sigma_{\text E}^2},\tau)
\end{equation}
\end{figure*}
\setcounter{equation}{\value{cnt3}}
\setcounter{equation}{50} 

{\noindent{where the expressions of ${{\cal I}_1}$, ${{\cal I}_2}$ and ${{\cal I}_3}$ are listed in (\ref{I_1_lemma}), (\ref{I_2_lemma}) and (\ref{I_3_lemma}), respectively. Moreover, the term $\chi$ in the expression of ${{\cal I}_1}$ is defined as }}
\begin{equation}
\label{Chi}
\begin{aligned} 
\chi& = \Psi(1,j_{\text B},0,\frac{1}{2\beta_1^2\sigma_{\text B}^2},0,\tau) \\
&= {\int\nolimits_{0}^{\tau}\ln (1+\gamma_{\text B}) {\gamma_{\text B}^{j_{\text B}}} \exp\left(-\frac{\gamma_{\text B}}{2{\beta_1^2}\sigma_{\text B}^2}\right)d\gamma_{\text B}}
\end{aligned} 
\end{equation}

{\textit {Proof:}} Please see Appendix A.

Asymptotically, when $\sigma_{\text E}^2\rightarrow \infty$, a closed-form expression of the lower bound of the average secrecy rate is given in Lemma 2.

{\textit {Lemma 2:}} When $\sigma_{\text E}^2\rightarrow \infty$, the lower bound of the average secrecy rate of the proposed FDA-DM system in FTR fading can be written in the following closed-form expression,
\begin{equation}
\label{R_s_Low_lemma}
\begin{aligned}
& \bar{R}_s^{\text {Low}}(\gamma_{\text B},\gamma_{\text E})   = \frac{m_{\text B}^{m_{\text B}}}{{\ln 2}\Gamma(m_{\text B})}\sum\limits_{j_{\text B}=0}^{\infty}\frac{K_{\text B}^{j_{\text B}}d_{j_{\text B}} }{{j_{\text B}}!{j_{\text B}}!(2{\beta_1^2}\sigma_{\text B}^2)^{{j_{\text B}} +1}} \\
 & ~~~\cdot [{ \cal S}(j_{\text B}+1,\frac{1}{2\beta_1^2\sigma_{\text B}^2}) - \chi] - \frac{\ln(1+\tau)}{\ln 2}[1-F_{\gamma_{\text B}}(\tau)]
\end{aligned}
 \end{equation}

{\textit {Proof:}} Please see Appendix B.

\subsection{Secrecy Outage Probability}
The secrecy outage probability is defined as the probability that the instantaneous secrecy rate $R_s$ is less than a target secrecy rate $R_0$, i.e.,
\begin{align}
\label{SOP_1}
{P_{\text {out}}} &
 = {\text{Pr}} \left\{R_s(\gamma_{\text B},\gamma_{\text E}) < R_0 \right\}\\
 \label{SOP_2}
& = {\text{Pr}} \left\{\log_2 \frac{1+\gamma_{\text B}}{1+\gamma_{\text E}} < R_0 \right\} 
\end{align}

{\textit {Lemma 3:}} The SOP of the proposed FDA-DM system in FTR fading can be obtained by (\ref{SOP_2_lemma}).

{\textit {Proof:}} Please see Appendix C.

\newcounter{cnt4}
\setcounter{cnt4}{\value{equation}}
\setcounter{equation}{54}
\begin{figure*}[hbp]
\begin{equation}
\label{SOP_2_lemma}
\begin{aligned}
{P_{\text {out}}}
& = \frac{m_{\text B}^{m_{\text B}}}{\Gamma(m_{\text B})} 
  \frac{m_{\text E}^{m_{\text E}}}{\Gamma(m_{\text E})} 
  \sum\limits_{j_{\text B}=0}^{\infty} \sum\limits_{j_{\text E}=0}^{\infty}
\frac{K_{\text B}^{j_{\text B}}d_{j_{\text B}}{K_{\text E}^{j_{\text E}}d_{j_{\text E}} } \tau}{{j_{\text B}}!{{j_{\text E}}!{j_{\text E}}!}(2\mu\sigma_{\text E}^2)^{{j_{\text E}} +1}} \left[ \Psi(0,j_{\text E},j_{\text E}+2,0,\frac{1}{2\mu\sigma_{\text E}^2},\tau) \right.\\
& ~~~ \left. - \sum\limits_{n=0}^{j_{\text B}}\frac{1}{n!} \exp\left(-\frac{2^{R_0}-1}{2{\beta_1^2}\sigma_{\text B}^2}\right)
\sum\limits_{k=0}^{n}  {n \choose k} \frac{2^{kR_0}(2^{R_0}-1)^{n-k}}{(2{\beta_1^2}\sigma_{\text B}^2)^n} \Psi(0,j_{\text E}+k,j_{\text E}+2,\frac{2^{R_0}}{2{\beta_1^2}\sigma_{\text B}^2},\frac{1}{2\mu\sigma_{\text E}^2},\tau)\right]\\
 \end{aligned}
\end{equation}
\end{figure*}
\setcounter{equation}{\value{cnt4}}
\setcounter{equation}{55}

%
%

When $\sigma_{\text E}^2\rightarrow \infty$, the upper bound of the SOP of the proposed FDA-DM system in FTR fading can be obtained by the following closed-form expression
\begin{align}
\label{SOP_Up1}
{P_{\text {out}}^{\text {Up}}} &= {\text{Pr}} \left\{\log_2 \frac{1+\gamma_{\text B}}{1+\gamma_{\text E}} < R_0 \right\}\bigg|_{\sigma_{\text E}^2\rightarrow \infty}  \\
\label{SOP_Up2}
& = {\text{Pr}} \left\{ {\gamma_{\text B}} < 2^{R_0}(1+\tau) -1\right\}  \\
\label{SOP_Up3}
& =F_{\gamma_{\text B}}\left(2^{R_0}({1+\tau}) -1\right) \\
\label{SOP_Up4}
  &= \frac{m_{\text {B}}^{m_{\text {B}}}}{\Gamma(m_{\text {B}})} \sum\limits_{j_{\text {B}}=0}^{\infty} \frac{K_{\text {B}}^{j_{\text {B}}}d_{j_{\text {B}}}}{{j_{\text {B}}}!{j_{\text {B}}}!}\Upsilon \left({j_{\text {B}}}+1,\frac{{2^{R_0}({1+\tau}) -1}}{2{\beta_1^2}\sigma_{\text {B}}^2}\right).
\end{align}

%

\section{Numerical Results}

In this section, Monte Carlo experiments are conducted to verify the theoretical analysis, where the ZF method \cite{Hu_Random_FDA_DM}, \cite{Qiu_AN_FDA_DM}, \cite{Qiu_MB_FDA_DM}, \cite{Hu_AN_MB}-\cite{Ji_Nakagami_FDA_DM2},  the SVD method \cite{Cheng_SVD_FDA_DM}, and the NoAN method \cite{Zeng_FTR_PLS}-\cite{Bilim_SER_PLS} are included. The detailed simulation parameters are listed in Table II.

\begin{figure}
\centering
\includegraphics[angle=0,width=0.48\textwidth]{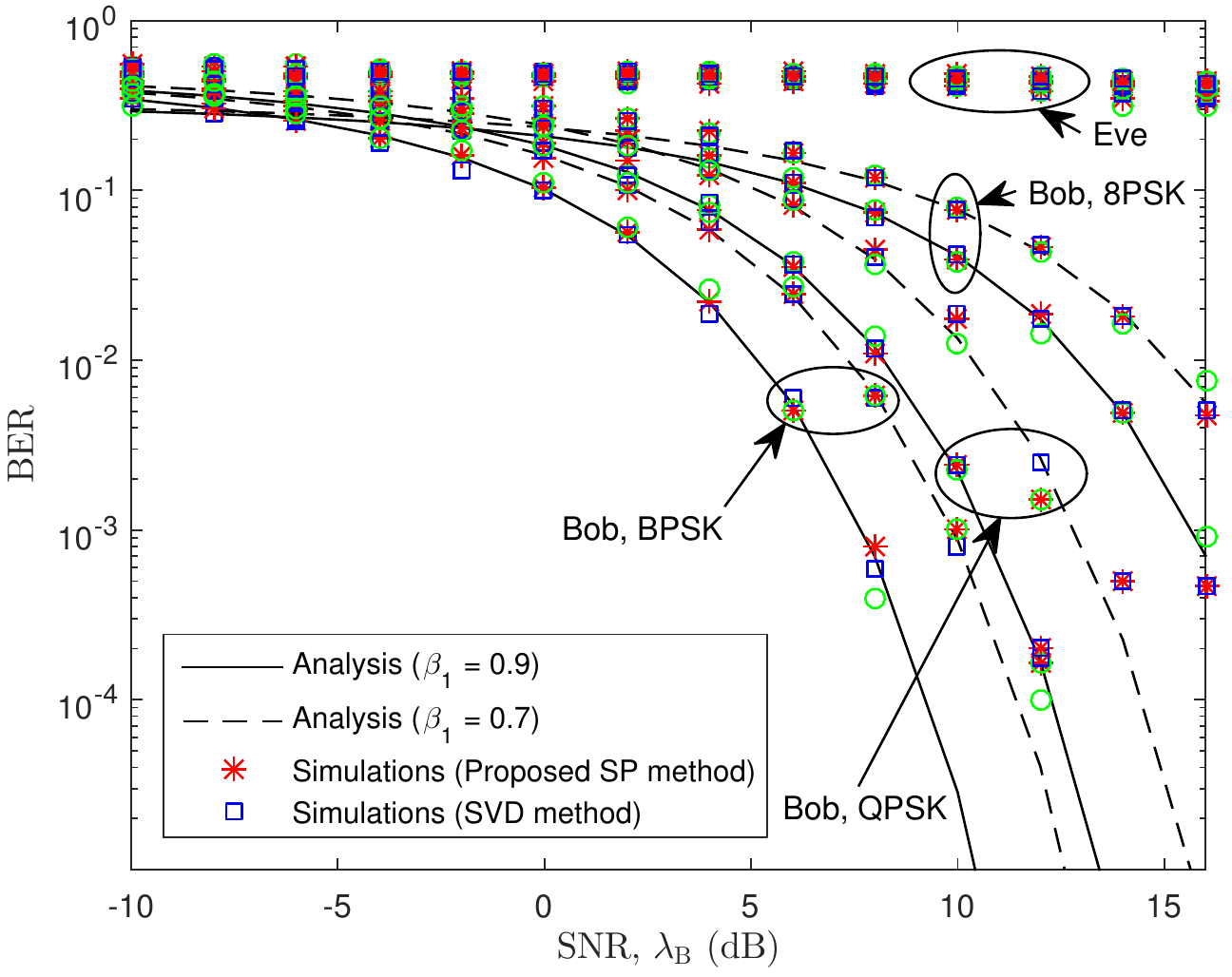}
\caption{BER versus SNR (dB) with $M$-PSK modulations. ($M=2,4,8$, $f_0=30$ GHz, $\Delta f=20$ kHz, $N=10$, $L=7$)}
\end{figure}

\begin{figure}
\centering
\includegraphics[angle=0,width=0.48\textwidth]{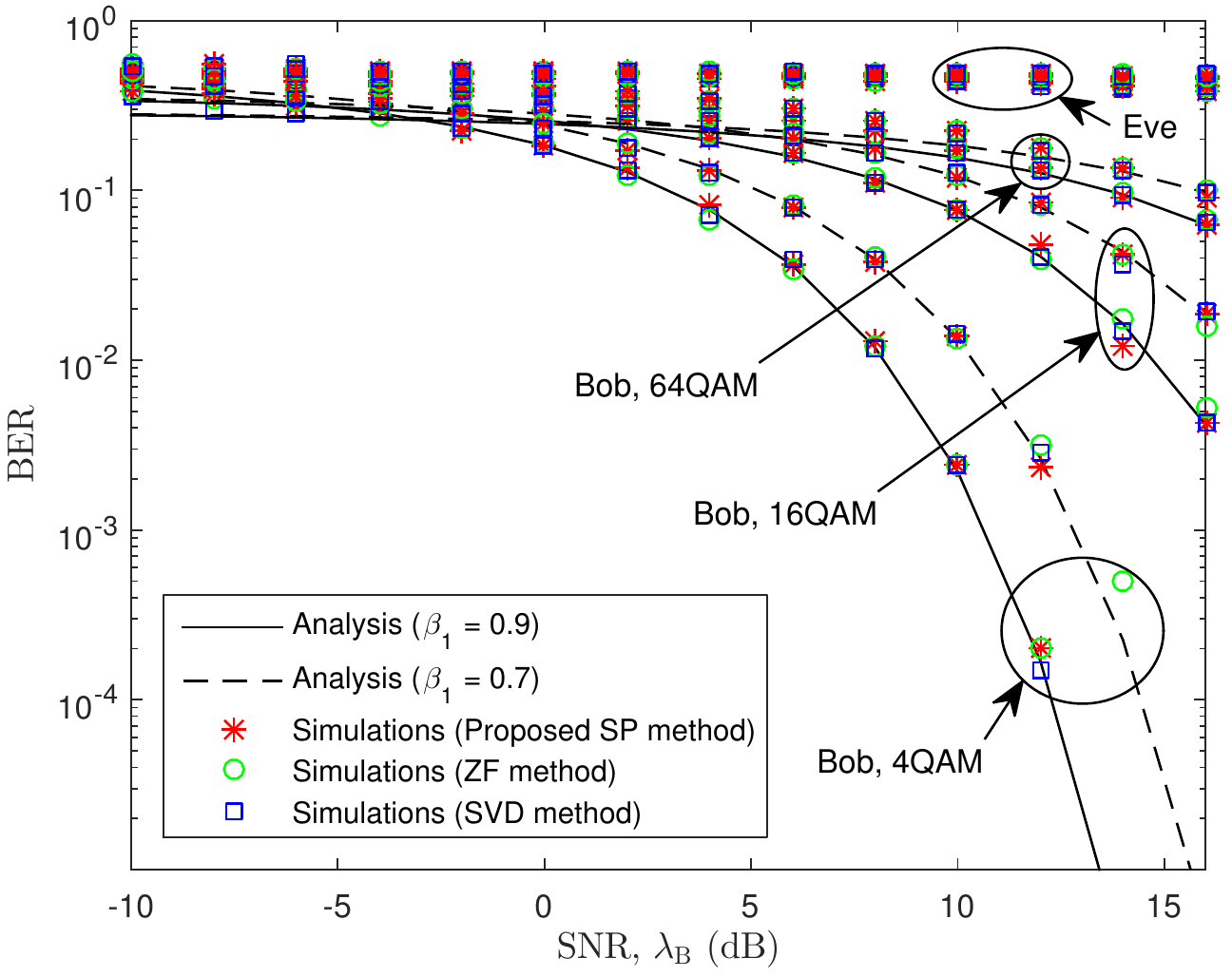}
\caption{BER versus SNR (dB) with $M$-QAM modulations. ($M=4,16,64$, $f_0=30$ GHz, $\Delta f=20$ kHz, $N=10$, $L=7$)}
\end{figure}

\begin{table}[!t]
\renewcommand{\arraystretch}{1.2}
\caption{Simulation Parameters}
\label{table2}
\centering
\begin{tabular}{ll}
\hline
{\bf Parameter} &{\bf Value } \\ [0.1ex]
\hline
Central frequency, ${f_0}$ & $30$ GHz  \\
Fixed frequency increment, ${\Delta f}$ & $20$ kHz \\
Number of FDA elements, $2N+1$ & $21$ \\
Number of subcarriers for each element, $L$ & $7$ \\
Total signal power, ${P_s}$ & $1$ \\
Power splitting factor, ${\beta_1}$ & $0.9,0.7$   \\
AWGN variance, $\delta_{\text B}^2$, $\delta_{\text E}^2$ & 1 \\
FTR parameters, $(m_{\text{B}},K_{\text{B}},\Delta_{\text{B}})$ & $(2.3,10,0.5)$\\
FTR parameters, $(m_{\text{E}},K_{\text{E}},\Delta_{\text{E}})$ & $(5.3,15,0.35)$\\
Location of Bob, $({r_{\text{B}}},{\theta_{\text{B}}},{\psi_{\text{B}}})$ & $(1~{\rm{km}},{20^ \circ },{30^ \circ })$\\
Location of Eve, $({r_{\text{E}}},{\theta_{\text{E}}},{\psi_{\text{E}}})$ & $(1.5~{\rm{km}},{-20^ \circ },{25^ \circ })$\\
Number of Monte Carlo experiments & $10^5$\\
Modulation mode & PSK, QAM  \\
\hline
\end{tabular}
\\
\end{table}


\subsection{Bit Error Rate}

Fig. 6 describes the BER performances with $\beta_1=0.9$ versus range, azimuth angle, and elevation angle, respectively, where QPSK modulation and SNR $=10$ dB are adopted. It can be clearly observed from Fig. 6 that only the receiver along the desired range, azimuth angle and elevation angle can achieve good BER performance, while the receivers at other locations cannot receive the confidential signal. In addition, compared with conventional FDA models \cite{Wang_FDA_Review}-\cite{Cheng_WFRFT_FDA_DM}, Fig. 6(c) also verifies that an extra dimension, i.e., elevation angle, is realized via the analytical model as analyzed in Section II.

Fig. 7 and Fig. 8 show the BER performances versus SNR (dB) with PSK and QAM modulations, respectively. It is observed that the proposed SP method can achieve almost the same BER performance as the conventional ZF and SVD methods. In addition, Bob exhibits a better BER performance while Eve's BER is much worse, which verifies the physical-layer security of the proposed FDA-DM model. Moreover,  the BER of Bob becomes better with larger $\beta_1$, which is because a  larger $\beta_1$ allocates more power to the useful signal.

\subsection{Secrecy Rate}

\begin{figure}
\centering
\includegraphics[angle=0,width=0.46\textwidth]{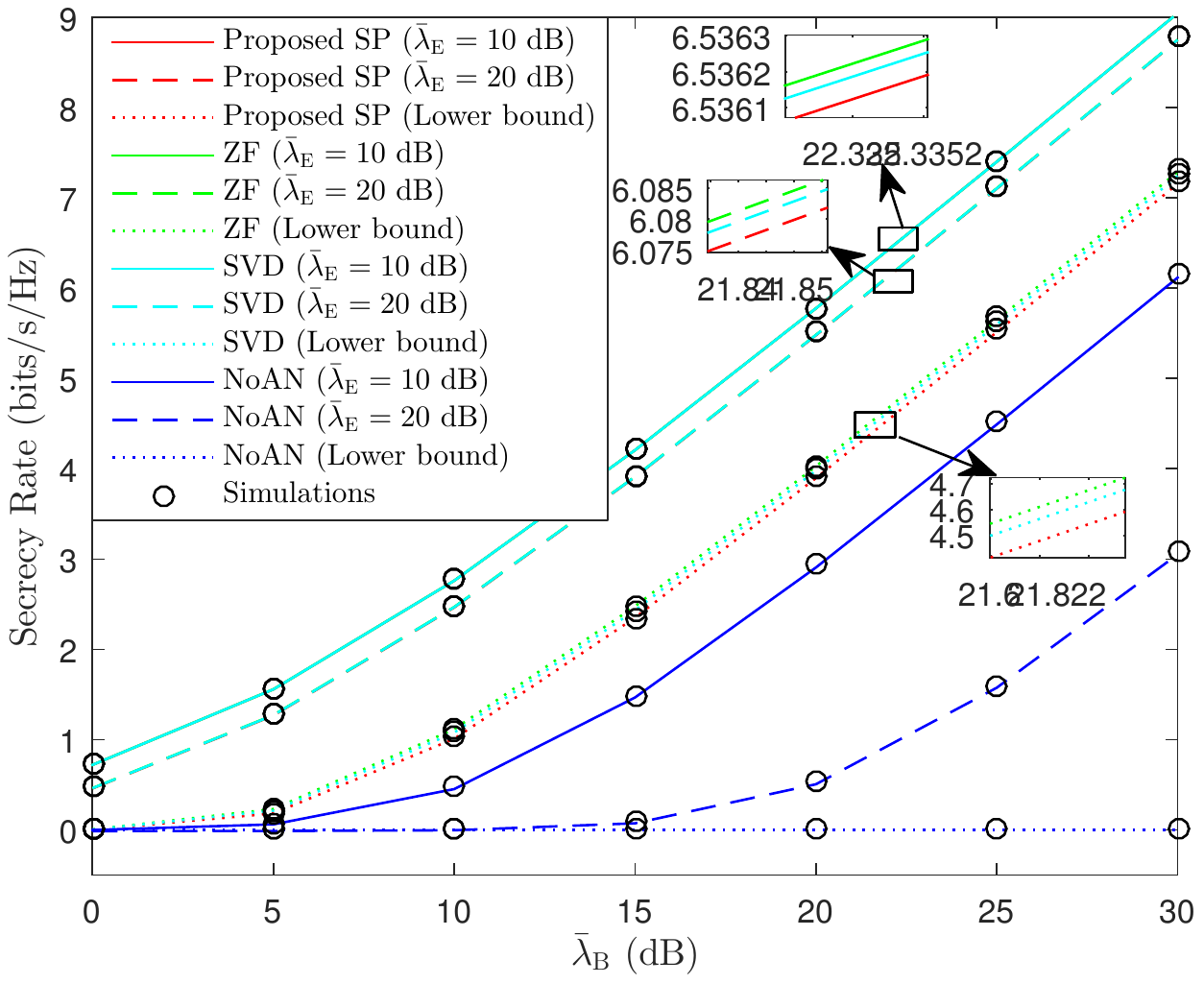}
\caption{Average secrecy rate $\bar{R}_s$ versus Bob's average SNR $\bar\lambda_{\text B}$ (dB) in FTR fading channels. ($f_0=30$ GHz, $\Delta f=20$ kHz, $N=10$, $L=7$, $\beta_1=0.9$)}
\end{figure}

\begin{figure}
\centering
\includegraphics[angle=0,width=0.46\textwidth]{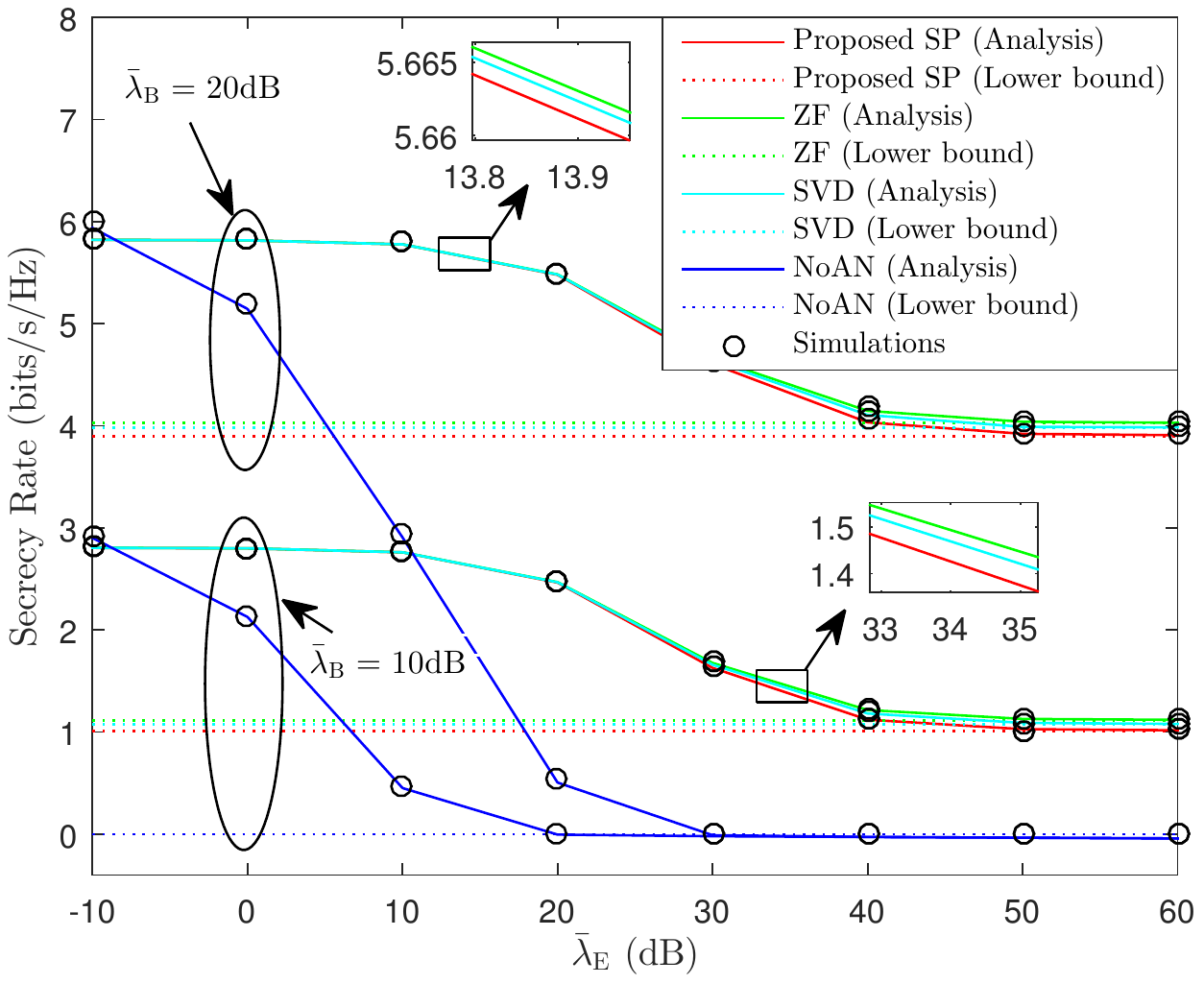}
\caption{Average secrecy rate $\bar{R}_s$ versus Eve's average SNR $\bar\lambda_{\text E}$ (dB) in FTR fading channels. ($f_0=30$ GHz, $\Delta f=20$ kHz, $N=10$, $L=7$, $\beta_1=0.9$)}
\end{figure}

In the simulations of the average SR, $\beta_1=0.9$ is adopted. Fig. 9 depicts the average SR of the proposed FDA-DM scheme versus $\bar\lambda_{\text B}$ in FTR fading channels, where the analytical results well match the simulated results with $10^5$ Monte Carlo experiments. First, it holds for all four PLS approaches that the average SR climbs with increasing $\bar\lambda_{\text B}$ when $\bar\lambda_{\text E}$ is fixed. Second, compared with ZF and SVD approaches, the proposed SP method can achieve almost the same average SR. Although there is actually a very small penalty on average SR, the proposed SP method is still much more competitive considering the considerable memory reduction. Moreover, compared with the NoAN approach, there is a positive lower bound for the average SR when $\bar\lambda_{\text E}\rightarrow \infty$, which means an absolutely positive average SR can always been realized as long as $\bar\lambda_{\text B}$ is big enough. By contrast, the average SR of NoAN approach reduces to zero when $\bar\lambda_{\text E}\rightarrow \infty$.

Fig. 10 illustrates the average SR of the proposed FDA-DM scheme versus $\bar\lambda_{\text E}$ in FTR fading channels. As expected, that the average SR decreases with increasing $\bar\lambda_{\text E}$ when $\bar\lambda_{\text B}$ is fixed. Differently, the average SRs of the proposed SP and the conventional ZF and SVD methods reduce to a positive lower bound, while the average SR of the NoAN method declines to zero. Similar to Fig. 9, almost the same average SR can be achieved for the proposed SP and the conventional ZF and SVD methods while the proposed SP distinguishes from the conventional ZF and SVD methods with much lower memory consumption as analyzed in Section III.A.

\subsection{Secrecy Outage Probability}

\begin{figure}
\centering
\includegraphics[angle=0,width=0.48\textwidth]{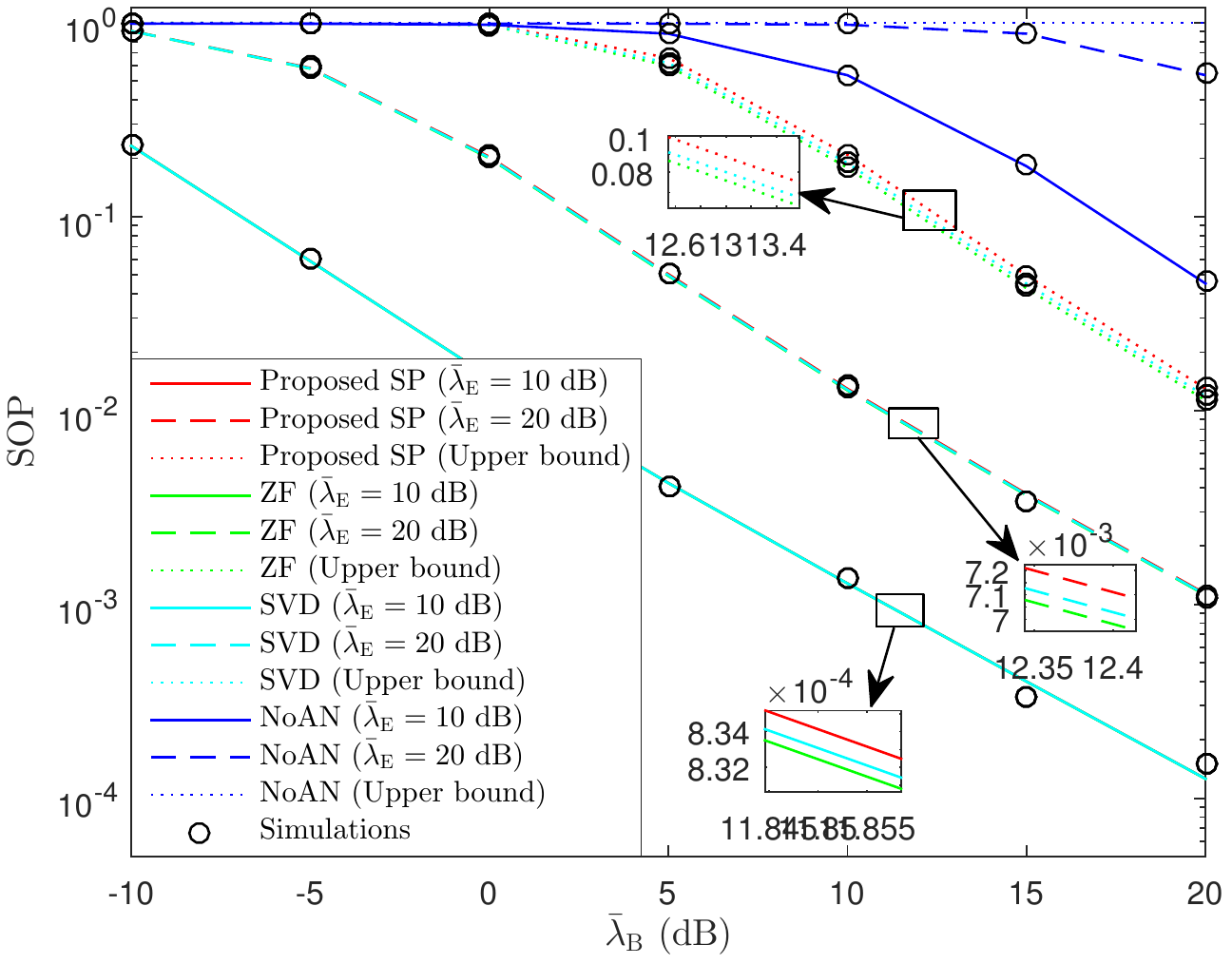}
\caption{SOP versus Bob's average SNR $\bar\lambda_{\text B}$ (dB) with $R_0=0$ in FTR fading channels. ($f_0=30$ GHz, $\Delta f=20$ kHz, $N=10$, $L=7$, $\beta_1=0.9$)}
\end{figure}

\begin{figure}
\centering
\includegraphics[angle=0,width=0.48\textwidth]{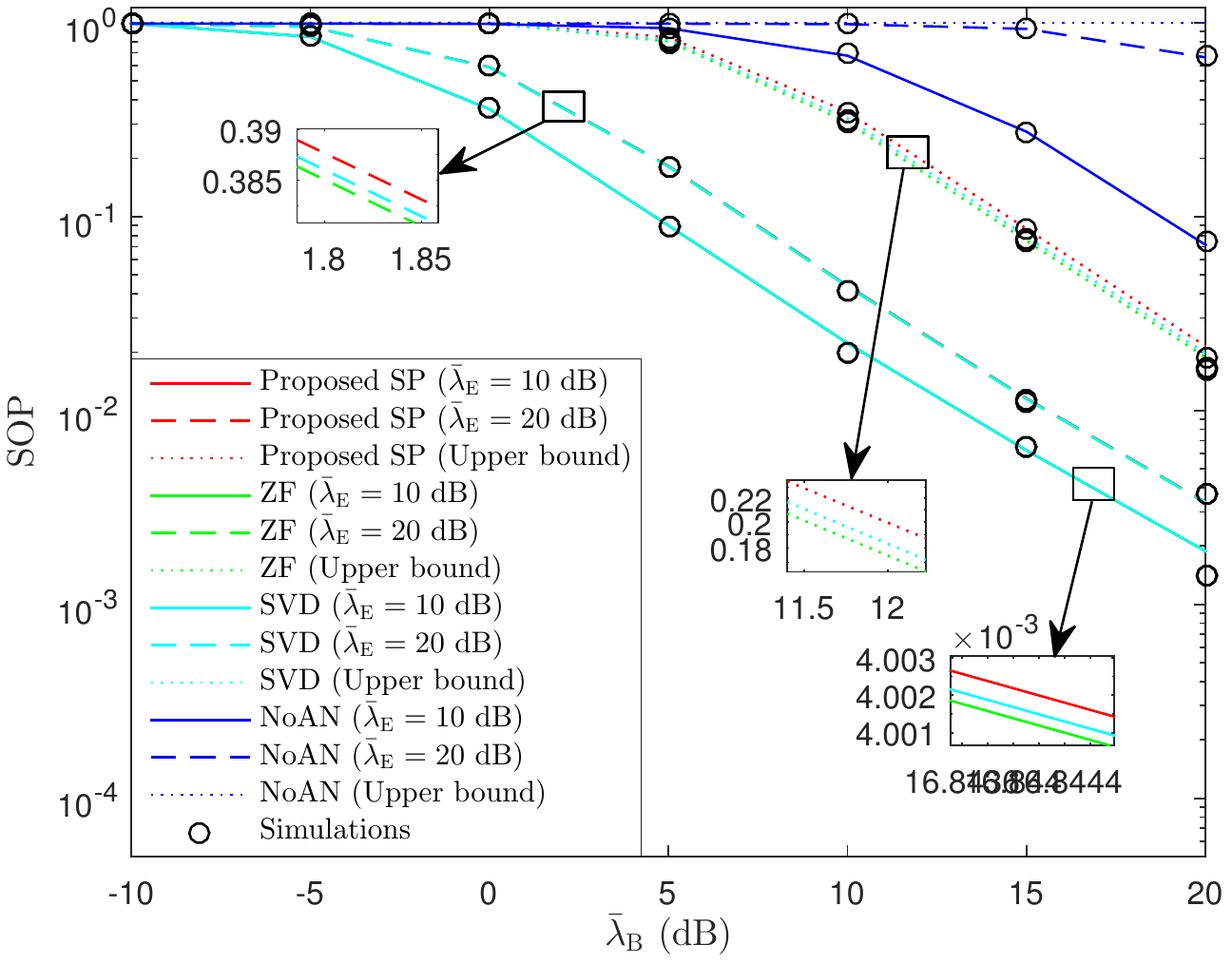}
\caption{SOP versus Bob's average SNR $\bar\lambda_{\text B}$ (dB) with $R_0=0.5$ bits/s/Hz in FTR fading channels. ($f_0=30$ GHz, $\Delta f=20$ kHz, $N=10$, $L=7$, $\beta_1=0.9$)}
\end{figure}

$\beta_1=0.9$ is adopted in the simulations of the SOP as well. Fig. 11 and Fig. 12 present the SOP of the proposed FDA-DM scheme versus $\bar\lambda_{\text B}$ in FTR fading channels with $R_0=0$ and $R_0=0.5$ bits/s/Hz, respectively. Comparing Fig. 11 and Fig. 12, it holds for all four PLS methods that the SOP decreases with smaller $R_0$. Given a specific $R_0$ and $\bar\lambda_{\text E}$, as expected, the SOP drops as $\bar\lambda_{\text B}$ increases. Compared with the conventional NoAN method, the proposed SP method can achieve much lower SOP. More importantly, there exists an upper bound of SOP  when $\bar\lambda_{\text E}\rightarrow \infty$ for the proposed SP method, while the SOP of the conventional NoAN method roars to 1. On the other hand, the proposed SP method can achieve almost the same SOP as the conventional ZF and SVD approaches with much lower memory consumption.

\begin{figure}
\centering
\includegraphics[angle=0,width=0.46\textwidth]{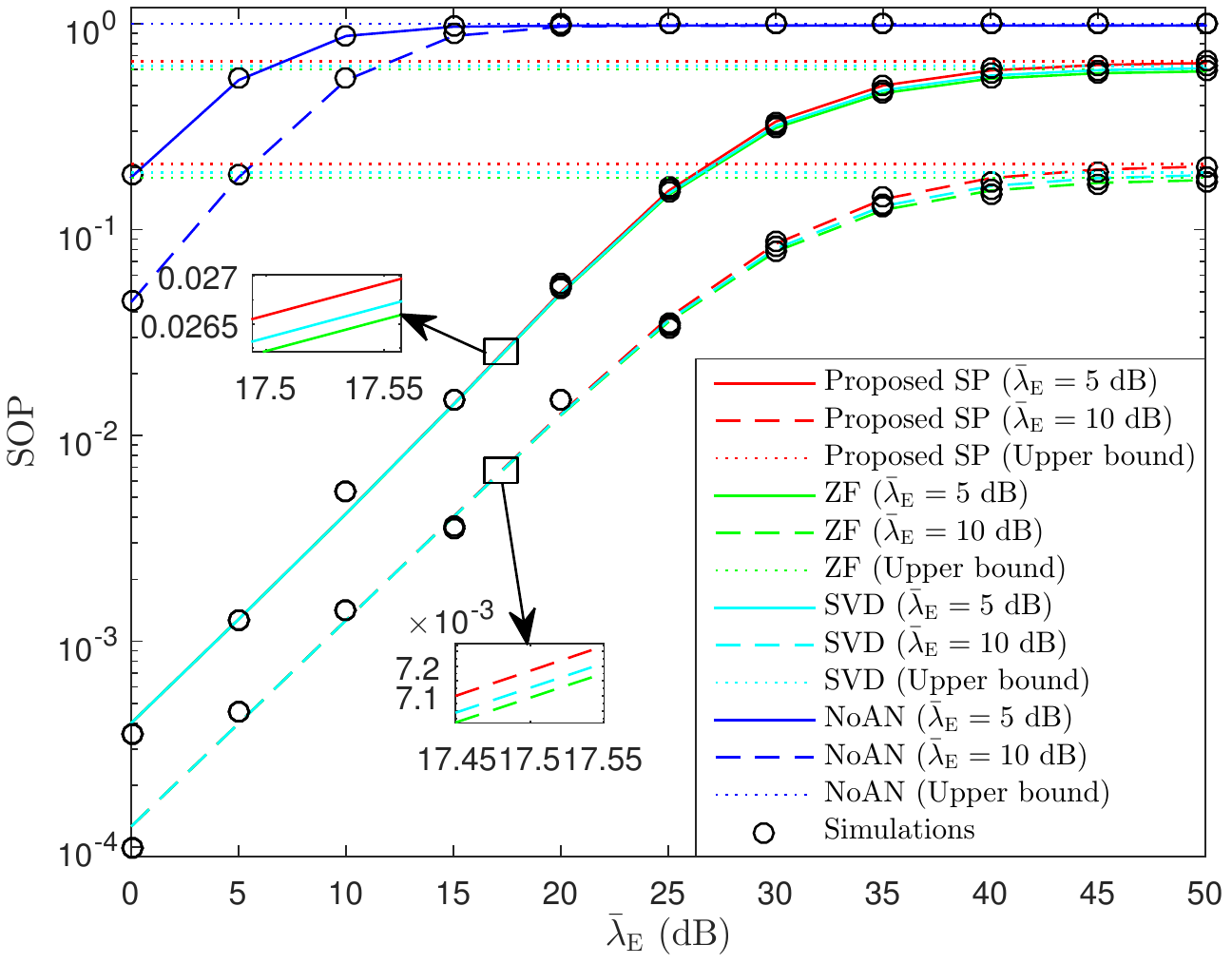}
\caption{SOP versus Eve's average SNR $\bar\lambda_{\text E}$ (dB) with $R_0=0$ in FTR fading channels. ($f_0=30$ GHz, $\Delta f=20$ kHz, $N=10$, $L=7$, $\beta_1=0.9$)}
\end{figure}

\begin{figure}
\centering
\includegraphics[angle=0,width=0.46\textwidth]{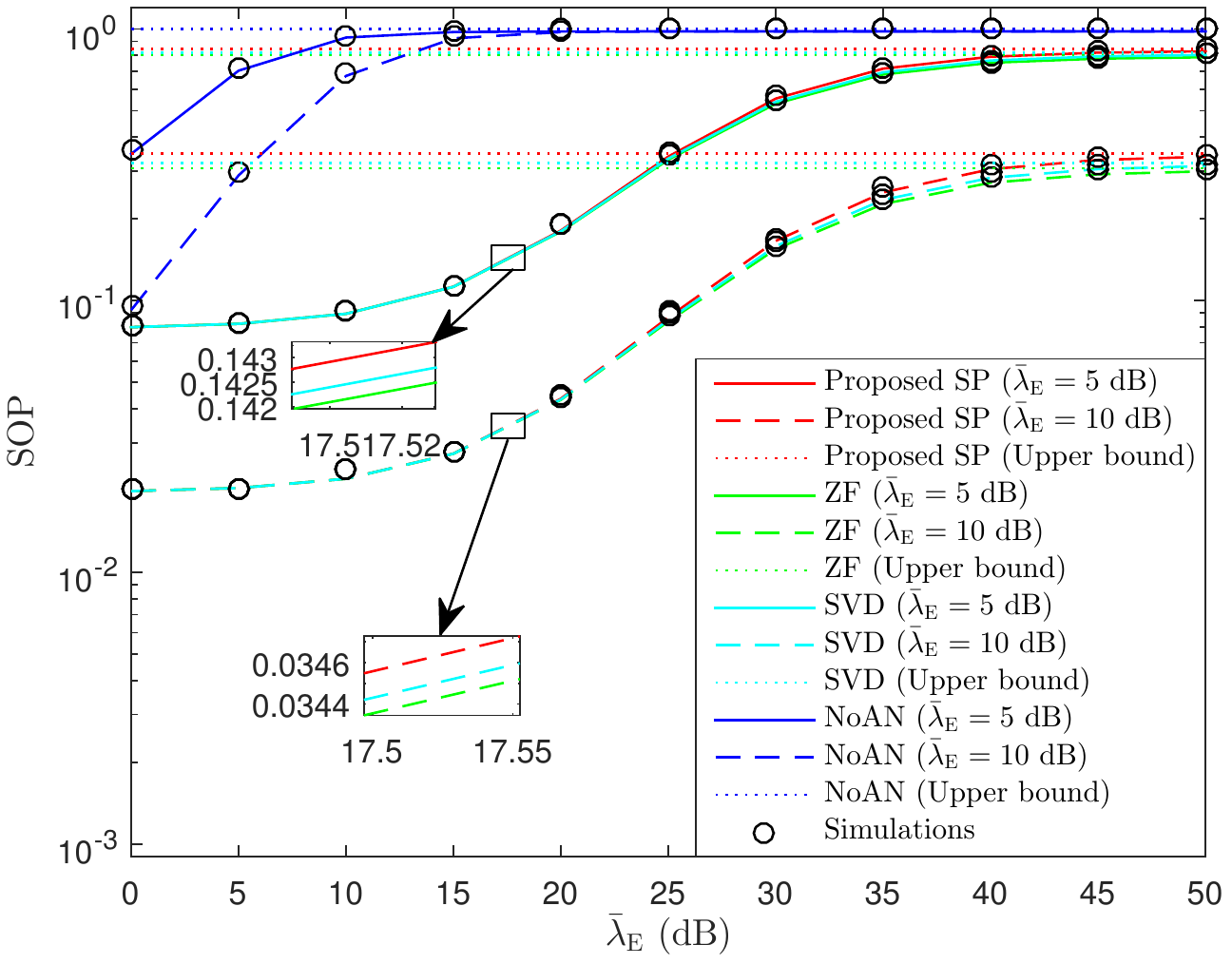}
\caption{SOP versus Eve's average SNR $\bar\lambda_{\text E}$ (dB) with $R_0=0.5$ bits/s/Hz in FTR fading channels. ($f_0=30$ GHz, $\Delta f=20$ kHz, $N=10$, $L=7$, $\beta_1=0.9$)}
\end{figure}

Fig. 13 and Fig. 14 illustrate the SOP of the proposed FDA-DM scheme versus $\bar\lambda_{\text E}$ in FTR fading channels with $R_0=0$ and $R_0=0.5$ bits/s/Hz, respectively. Analogous to Fig. 11 and Fig. 12, it can also be observed that a smaller $R_0$ produces a smaller SOP. With a specific $R_0$, the SOPs of the proposed SP method and the conventional ZF and SVD methods increase to an upper bound along with increasing $\bar\lambda_{\text E}$. But for the conventional NoAN method, it roars rapidly to 1 when $\bar\lambda_{\text E}$ increases. The advantage of the proposed SP method is verified again that much lower memory consumption yet achieves almost the same SOP as the ZF and SVD methods.

\newcounter{cnt5}
\setcounter{cnt5}{\value{equation}}
\setcounter{equation}{59}
\begin{figure*}[hbp]
\hrulefill
\begin{align}
\label{I_1_appendix1}
{{\cal I}_1} 
& = {\frac{1}{\ln 2}\int\nolimits_{0}^{\tau}\ln(1+\gamma_{\text B})f_{\gamma_{\text B}}(\gamma_{\text B})F_{\gamma_{\text E}}(\gamma_{\text B})d\gamma_{\text B} } + {\frac{1}{\ln 2}\int\nolimits_{\tau}^{\infty}\ln(1+\gamma_{\text B})f_{\gamma_{\text B}}(\gamma_{\text B})F_{\gamma_{\text E}}(\gamma_{\text B})d\gamma_{\text B} }\\
\label{I_1_appendix2}
& = \frac{m_{\text B}^{m_{\text B}}{m_{\text E}^{m_{\text E}}}}{{\ln 2}\Gamma(m_{\text B}){\Gamma(m_{\text E})}} 
\sum\limits_{j_{\text B}=0}^{\infty}\sum\limits_{j_{\text E}=0}^{\infty}
\frac{K_{\text B}^{j_{\text B}}d_{j_{\text B}} {K_{\text E}^{j_{\text E}}d_{j_{\text E}}}}{{j_{\text B}}!{j_{\text B}}!{{j_{\text E}}!{j_{\text E}}!} {(2\beta_1^2\sigma_{\text B}^2)^{{j_{\text B}} +1}} } \underbrace{ \int\nolimits_{0}^{\tau} \ln(1+\gamma_{\text B})  
 {{\gamma_{\text B}^{j_{\text B}}}} \exp\left(-\frac{\gamma_{\text B}}{2\beta_1^2\sigma_{\text B}^2}\right)
   \Upsilon \left({j_{\text E}}+1,\frac{\gamma_{\text B}}{ {2\sigma_{\text E}^2}(\eta-\mu \gamma_{\text B})}\right) d\gamma_{\text B} }_ {{\cal I}_{1,1}}\\
   \label{I_1_appendix3}
&~~~+ \frac{m_{\text B}^{m_{\text B}}}{{\ln 2}\Gamma(m_{\text B})} 
\sum\limits_{j_{\text B}=0}^{\infty}\frac{K_{\text B}^{j_{\text B}}d_{j_{\text B}} }{{j_{\text B}}!{j_{\text B}}!{(2\beta_1^2\sigma_{\text B}^2)^{{j_{\text B}} +1}}}
\underbrace {\int\nolimits_{\tau}^{\infty} \ln(1+\gamma_{\text B})  
 {{\gamma_{\text B}^{j_{\text B}}}} \exp\left(-\frac{\gamma_{\text B}}{2\beta_1^2\sigma_{\text B}^2}\right)d\gamma_{\text B}}_{{\cal I}_{1,2}}
\end{align}
\end{figure*}
\setcounter{equation}{\value{cnt5}}
\setcounter{equation}{62}

\section{Conclusion}

In this paper, we presented a low-memory-consumption single-point AN-aided secure DM transmission scheme for 5G and beyond communications based on symmetrical multi-carrier FDA, which significantly outperforms the conventional ZF and SVD approaches with only a very small penalty on secrecy rate and secrecy outage probability. In the proposed FDA-DM scheme, the FDA was analyzed in three dimensions, i.e., range, azimuth angle, and elevation angle. Moreover, the secrecy rate and secrecy outage probability of the proposed FDA-DM scheme were analyzed, for the first time, in FTR fading channels, which provide a better fit for small-scale fading measurements in mmWave communications. The closed-form expressions of lower SR bound and upper SOP bound were derived and numerical demonstrations by Monte Carlo experiments were provided as well. One future work is to investigate the PLS performance of the proposed low-memory-consumption FDA-DM scheme with multiple legitimate users.

\section*{Appendix}

\subsection{Proof of Lemma 1}

Here, we derive the expression of ${{\cal I}_1}$. Substituting (\ref{pdf_gamma_B}) and  (\ref{cdf_gamma_E4}) into (\ref{average_R_s}), we can obtain (\ref{I_1_appendix1})-(\ref{I_1_appendix3}), where $\Upsilon \left({j_{\text E}}+1,\frac{\gamma_{\text B}}{ {2\sigma_{\text E}^2}(\eta-\mu \gamma_{\text B})}\right)$ can be written as [39, Eq. (8.354.1)]
\begin{equation}
\label{Upsilon_I_1}
\begin{aligned}
& \Upsilon \left({j_{\text E}}+1,\frac{\gamma_{\text B}}{ {2\sigma_{\text E}^2}(\eta-\mu \gamma_{\text B})}\right)
= \\
&  {j_{\text E}}! \left(1-\exp\left(- \frac{\gamma_{\text B}}{ {2\sigma_{\text E}^2}(\eta-\mu \gamma_{\text B})}\right) \sum\limits_{n=0}^{j_{\text E}} \frac{1}{n!}\left(\frac{\gamma_{\text B}}{ {2\sigma_{\text E}^2}(\eta-\mu \gamma_{\text B})}\right)^{n}\right)
\end{aligned}
\end{equation} 


Substituting (\ref{Upsilon_I_1})  into (\ref{I_1_appendix2}) and using (\ref{Psi_v_tau}), we can acquire
\begin{equation}
\label{I_1_1}
\begin{aligned}
{\cal I}_{1,1} 
   & = {j_{\text E}}!\Psi(1,j_{\text B},0,\frac{1}{2\beta_1^2\sigma_{\text B}^2},0,\tau)  - \\
   & ~~~ {j_{\text E}}! \sum\limits_{n=0}^{j_{\text E}} \frac{1}{n!({2\mu\sigma_{\text E}^2})^n}
    \Psi(1,j_{\text B}+n,n,\frac{1}{2\beta_1^2\sigma_{\text B}^2},\frac{1}{2\mu\sigma_{\text E}^2},\tau)
\end{aligned}
\end{equation} 

Similarly, ${\cal I}_{1,2}$ can be calculated by 
\begin{equation}
\label{I_1_2}
\begin{aligned}
{\cal I}_{1,2}  = \Psi(1,j_{\text B},0,\frac{1}{2\beta_1^2\sigma_{\text B}^2},0,\infty)
 - \Psi(1,j_{\text B},0,\frac{1}{2\beta_1^2\sigma_{\text B}^2},0,\tau) 
\end{aligned}
\end{equation} 

Substituting (\ref{I_1_1}) and (\ref{I_1_2}) into (\ref{I_1_appendix2}) and (\ref{I_1_appendix3}), we can get the expression of ${{\cal I}_1}$ in (\ref{I_1_lemma}).  ${{\cal I}_2}$ and ${{\cal I}_3}$ can be derived in the same way, which ends the proof of Lemma 1.

\subsection{Proof of Lemma 2}

Observing (\ref{gamma_E2}), when $\sigma_{\text E}^2 \rightarrow \infty$, we can subsequently obtain $\lambda_{\text E} \rightarrow \infty$ and  $\gamma_{\text E} \rightarrow \tau$. Therefore, $\bar{R}_s^{\text {Low}}$ can be written as
\begin{equation}
\label{R_s_Low_appendix1}
\begin{aligned}
  \bar{R}_s^{\text {Low}}(\gamma_{\text B},\gamma_{\text E})  & =  \frac{1}{\ln 2}\int\nolimits_{0}^{\infty}\left[\ln (1+\gamma_{\text B}) - \ln(1+\tau)\right]^{+}f_{\gamma_{\text B}}(\gamma_{\text B})d\gamma_{\text B}\\
& = \frac{1}{\ln 2} \int\nolimits_{\tau}^{\infty}\ln (1+\gamma_{\text B})f_{\gamma_{\text B}}(\gamma_{\text B})d\gamma_{\text B} \\
& ~~~ -  \frac{1}{\ln 2}\int\nolimits_{\tau}^{\infty}\ln(1+\tau)f_{\gamma_{\text B}}(\gamma_{\text B})d\gamma_{\text B}
\end{aligned}
\end{equation} 

Replacing (\ref{pdf_lambda_i}) and (\ref{pdf_gamma_B}) into (\ref{R_s_Low_appendix1}) yields (\ref{R_s_Low_appendix21}) and (\ref{R_s_Low_appendix22}), and  substituting (\ref{Psi_v_tau}) into (\ref{R_s_Low_appendix22}) ends the proof of Lemma 2.

\newcounter{cnt6}
\setcounter{cnt6}{\value{equation}}
\setcounter{equation}{66}
\begin{figure*}[hbp]
\begin{align}
\label{R_s_Low_appendix21}
 \bar{R}_s^{\text {Low}}(\gamma_{\text B},\gamma_{\text E})
& =  \frac{1}{\ln 2}\int\nolimits_{\tau}^{\infty}\ln (1+\gamma_{\text B})\frac{m_{\text B}^{m_{\text B}}}{\Gamma(m_{\text B})} \sum\limits_{j_{\text B}=0}^{\infty}\frac{K_{\text B}^{j_{\text B}}d_{j_{\text B}} }{{j_{\text B}}!{j_{\text B}}!}\frac{\gamma_{\text B}^{j_{\text B}}}{(2{\beta_1^2}\sigma_{\text B}^2)^{{j_{\text B}} +1}} \exp\left(-\frac{\gamma_{\text B}}{2{\beta_1^2}\sigma_{\text B}^2}\right)d\gamma_{\text B} -  \frac{\ln(1+\tau)}{\ln 2}[1-F_{\gamma_{\text B}}(\tau)]\\
\label{R_s_Low_appendix22}
& =\frac{m_{\text B}^{m_{\text B}}}{{\ln 2}\Gamma(m_{\text B})}\sum\limits_{j_{\text B}=0}^{\infty}\frac{K_{\text B}^{j_{\text B}}d_{j_{\text B}} }{{j_{\text B}}!{j_{\text B}}!(2{\beta_1^2}\sigma_{\text B}^2)^{{j_{\text B}} +1}} {\int\nolimits_{\tau}^{\infty}\ln (1+\gamma_{\text B}) {\gamma_{\text B}^{j_{\text B}}} \exp\left(-\frac{\gamma_{\text B}}{2{\beta_1^2}\sigma_{\text B}^2}\right)d\gamma_{\text B}} - \frac{\ln(1+\tau)}{\ln 2}[1-F_{\gamma_{\text B}}(\tau)]
\end{align}
\end{figure*}
\setcounter{equation}{\value{cnt6}}
\setcounter{equation}{68}

\subsection{Proof of Lemma 3}
Given a fixed secrecy rate $R_0$, the SOP in (\ref{SOP_2}) can be further written as
\begin{align}
\label{SOP_appendix1}
{P_{\text {out}}} 
  & = {\text{Pr}} \left\{ {\gamma_{\text B}} < 2^{R_0}({1+\gamma_{\text E}}) -1\right\} \\
  \label{SOP_appendix2}
  & = \int\nolimits_{0}^{\infty} F_{\gamma_{\text B}}\left(2^{R_0}({1+\gamma_{\text E}}) -1\right)
 f_{\gamma_{\text E}}({\gamma_{\text E}})d{\gamma_{\text E}}
 \end{align}

\newcounter{cnt7}
\setcounter{cnt7}{\value{equation}}
\setcounter{equation}{70}
\begin{figure*}[hbp]
\hrulefill
\begin{align}
\label{SOP_appendix21}
{P_{\text {out}}} 
 & = \int\nolimits_{0}^{\tau} 
 \frac{m_{\text B}^{m_{\text B}}}{\Gamma(m_{\text B})} \sum\limits_{j_{\text B}=0}^{\infty} \frac{K_{\text B}^{j_{\text B}}d_{j_{\text B}}}{{j_{\text B}}!{j_{\text B}}!}\Upsilon \left({j_{\text B}}+1,\frac{{2^{R_0}({1+\gamma_{\text E}}) -1}}{2{\beta_1^2}\sigma_{\text B}^2}\right)  \frac{\eta}{(\eta-\mu \gamma_{\text E})^2} 
\frac{m_{\text E}^{m_{\text E}}}{\Gamma(m_{\text E})} \sum\limits_{j_{\text E}=0}^{\infty}\frac{K_{\text E}^{j_{\text E}}d_{j_{\text E}} }{{j_{\text E}}!{j_{\text E}}!}\frac{(\frac{\gamma_{\text E}}{{\eta} - \mu \gamma_{\text E}})^{j_{\text E}}}{(2\sigma_{\text E}^2)^{{j_{\text E}} +1}} \exp\left(-\frac{\frac{\gamma_{\text E}}{{\eta} - \mu \gamma_{\text E}}}{2\sigma_{\text E}^2}\right) d{\gamma_{\text E}}\\
\label{SOP_appendix22}
& =   
  \frac{m_{\text B}^{m_{\text B}}}{\Gamma(m_{\text B})} 
  \frac{m_{\text E}^{m_{\text E}}}{\Gamma(m_{\text E})} 
  \sum\limits_{j_{\text B}=0}^{\infty} \sum\limits_{j_{\text E}=0}^{\infty}
\frac{K_{\text B}^{j_{\text B}}d_{j_{\text B}}{K_{\text E}^{j_{\text E}}d_{j_{\text E}} } \tau}{{j_{\text B}}!{j_{\text B}}!{{j_{\text E}}!{j_{\text E}}!}(2\mu\sigma_{\text E}^2)^{{j_{\text E}} +1}}\\
\label{SOP_appendix23}
& ~~~ \cdot
  \underbrace{\int\nolimits_{0}^{\tau} 
 \Upsilon \left({j_{\text B}}+1,\frac{{2^{R_0}({1+\gamma_{\text E}}) -1}}{2{\beta_1^2}\sigma_{\text B}^2}\right)   
\frac{\gamma_{\text E}^{j_{\text E}}}{ ({{\tau} -  \gamma_{\text E}})^{j_{\text E}+2} } 
\exp\left(-\frac{{\gamma_{\text E}}}{2\mu\sigma_{\text E}^2 ({{\tau} -  \gamma_{\text E}})}\right) d{\gamma_{\text E}}}_{{\cal A}}
 \end{align}
\end{figure*}
\setcounter{equation}{\value{cnt7}}
\setcounter{equation}{73}

\newcounter{cnt8}
\setcounter{cnt8}{\value{equation}}
\setcounter{equation}{73}
\begin{figure*}[hbp]
 \begin{align}
  \label{A_appendix1}
{{\cal A}}
& = \int\nolimits_{0}^{\tau} 
j_{\text B}!\left(1 - \exp\left(- \frac{{2^{R_0}({1+\gamma_{\text E}}) -1}}{2{\beta_1^2}\sigma_{\text B}^2}\right)
\sum\limits_{n=0}^{j_{\text B}} \frac{1}{n!} 
\left(  \frac{{2^{R_0}({1+\gamma_{\text E}}) -1}}{2{\beta_1^2}\sigma_{\text B}^2}\right)^n\right)\frac{\gamma_{\text E}^{j_{\text E}}}{ ({{\tau} -  \gamma_{\text E}})^{j_{\text E}+2} } 
\exp\left(-\frac{{\gamma_{\text E}}}{2\mu\sigma_{\text E}^2 ({{\tau} -  \gamma_{\text E}})}\right) d{\gamma_{\text E}}\\
  \label{A_appendix2}
& = j_{\text B}! \int\nolimits_{0}^{\tau} \frac{\gamma_{\text E}^{j_{\text E}}}{ ({{\tau} -  \gamma_{\text E}})^{j_{\text E}+2} } 
\exp\left(-\frac{{\gamma_{\text E}}}{2\mu\sigma_{\text E}^2 ({{\tau} -  \gamma_{\text E}})}\right) d{\gamma_{\text E}}\\
  \label{A_appendix3}
& ~~~ -  \sum\limits_{n=0}^{j_{\text B}}\frac{j_{\text B}!}{n!} \underbrace{ \int\nolimits_{0}^{\tau}
\frac{\gamma_{\text E}^{j_{\text E}}}{ ({{\tau} -  \gamma_{\text E}})^{j_{\text E}+2} } 
\left(  \frac{{2^{R_0}({1+\gamma_{\text E}}) -1}}{2{\beta_1^2}\sigma_{\text B}^2}\right)^n
\exp\left(-\frac{{\gamma_{\text E}}}{2\mu\sigma_{\text E}^2 ({{\tau} -  \gamma_{\text E}})} - \frac{{2^{R_0}({1+\gamma_{\text E}}) -1}}{2{\beta_1^2}\sigma_{\text B}^2}\right)
d{\gamma_{\text E}} }_{{\cal B}}\\
  \label{A_appendix4}
& = j_{\text B}! \Psi(0,j_{\text E},j_{\text E}+2,0,\frac{1}{2\mu\sigma_{\text E}^2},\tau) - \sum\limits_{n=0}^{j_{\text B}}\frac{j_{\text B}!}{n!} {\cal B}
 \end{align}
\end{figure*}
\setcounter{equation}{\value{cnt8}}
\setcounter{equation}{77}

Then, replacing (\ref{pdf_lambda_i}), (\ref{pdf_gamma_B}) and (\ref{pdf_gamma_E2}) into (\ref{SOP_appendix2}) yields (\ref{SOP_appendix21})-(\ref{SOP_appendix23}). Using [39, Eq. (8.354.1)], (\ref{SOP_appendix23}) can be calculated by (\ref{A_appendix1})-(\ref{A_appendix4}). According to the binomial theorem [39, Eq. (1.111)], we can write 
 \begin{align}
  \label{Binary_Expend1}
 \left(  \frac{{2^{R_0}({1+\gamma_{\text E}}) -1}}{2{\beta_1^2}\sigma_{\text B}^2}\right)^n 
  & =  \left(  \frac{2^{R_0}}{2{\beta_1^2}\sigma_{\text B}^2} \gamma_{\text E} + \frac{2^{R_0} -1 }{2{\beta_1^2}\sigma_{\text B}^2}\right)^n 
  \\
    \label{Binary_Expend2}
  & = \frac{2^{nR_0}}{(2{\beta_1^2}\sigma_{\text B}^2)^n}
 \sum\limits_{k=0}^{n}  {n \choose k}
 \gamma_{\text E} ^k \left(\frac{2^{R_0} -1 }{2^{R_0}}\right)^{n-k}
  \end{align}

Therefore, ${{\cal B}}$ can be acquired in (\ref{B_appendix1})-(\ref{B_appendix3}) by replacing (\ref{Binary_Expend2}) into (\ref{A_appendix3}). Finally, substituting (\ref{A_appendix4}) and (\ref{B_appendix3}) into (\ref{SOP_appendix23}) ends the proof of Lemma 3.

\newcounter{cnt9}
\setcounter{cnt9}{\value{equation}}
\setcounter{equation}{79}
\begin{figure*}[hbp]
\hrulefill
 \begin{align}
  \label{B_appendix1}
{{\cal B}}
& =  \int\nolimits_{0}^{\tau}
\frac{\gamma_{\text E}^{j_{\text E}}}{ ({{\tau} -  \gamma_{\text E}})^{j_{\text E}+2} } 
\frac{2^{nR_0}}{(2{\beta_1^2}\sigma_{\text B}^2)^n}
 \sum\limits_{k=0}^{n}  (_k^n)
 \gamma_{\text E} ^k \left(\frac{2^{R_0} -1 }{2^{R_0}}\right)^{n-k}
\exp\left(-\frac{{\gamma_{\text E}}}{2\mu\sigma_{\text E}^2 ({{\tau} -  \gamma_{\text E}})} - \frac{{2^{R_0}({1+\gamma_{\text E}}) -1}}{2{\beta_1^2}\sigma_{\text B}^2}\right)
d{\gamma_{\text E}} \\
  \label{B_appendix2}
& = \exp\left(-\frac{2^{R_0}-1}{2{\beta_1^2}\sigma_{\text B}^2}\right)
\sum\limits_{k=0}^{n}  (_k^n) \frac{2^{kR_0}(2^{R_0}-1)^{n-k}}{(2{\beta_1^2}\sigma_{\text B}^2)^n}
\int\nolimits_{0}^{\tau}
\frac{\gamma_{\text E}^{j_{\text E} + k }}{ ({{\tau} -  \gamma_{\text E}})^{j_{\text E}+2} } 
\exp\left(-\frac{{\gamma_{\text E}}}{2\mu\sigma_{\text E}^2 ({{\tau} -  \gamma_{\text E}})} - \frac{2^{R_0}\gamma_{\text E}}{2{\beta_1^2}\sigma_{\text B}^2}\right)
d{\gamma_{\text E}} \\
  \label{B_appendix3}
& = \exp\left(-\frac{2^{R_0}-1}{2{\beta_1^2}\sigma_{\text B}^2}\right)
\sum\limits_{k=0}^{n}  (_k^n) \frac{2^{kR_0}(2^{R_0}-1)^{n-k}}{(2{\beta_1^2}\sigma_{\text B}^2)^n} \Psi(0,j_{\text E}+k,j_{\text E}+2,\frac{2^{R_0}}{2{\beta_1^2}\sigma_{\text B}^2},\frac{1}{2\mu\sigma_{\text E}^2},\tau)
 \end{align}
\end{figure*}
\setcounter{equation}{\value{cnt9}}
\setcounter{equation}{82}


\section*{Acknowledgment}

The authors would like to thank Dr. Jiakang Zheng and Prof. Jiayi Zhang who provided constructive suggestions about the theoretical PDF calculation of FTR fading channels.

\begin{IEEEbiography}[{\includegraphics[width=1in,height=1.25in,clip,keepaspectratio]{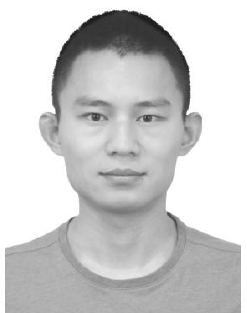}}]{Qian Cheng}
received the B.S. degree in Information and Communication Engineering from Xidian University, Xi'an, China, in 2014, and the M.S. degree in Information and Communication Engineering from the National University of Defense Technology (NUDT), Changsha, China, in 2016, where he is currently pursuing the Ph.D. degree.

 He is also a Visiting Ph.D. Researcher with the Institute of Electronics, Communications and Information Technology (ECIT), Queen's University Belfast, Belfast, U.K. His  research interests include physical layer security and directional modulation.
\end{IEEEbiography}

\begin{IEEEbiography}[{\includegraphics[width=1in,height=1.25in,clip,keepaspectratio]{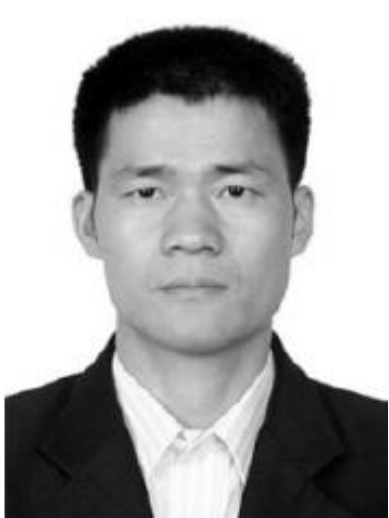}}]{Shilian Wang}
received his B.S. and Ph.D. degrees in information and communication engineering from National University of Defense Technology (NUDT), Changsha, China, in 1998 and 2004, respectively. Since 2004, he continued research in wireless communications at NUDT, where he later became a Professor. From 2008 to 2009, he was a visiting scholar with the Department of Electronic and Electrical Engineering at Columbia University (CU), New York, USA.

He is currently the Head of the Laboratory of Advanced Communication Technology at the School of Electronic Science, NUDT. He has authored or co-authored two books, 26 journal papers, and 20 conference papers. His research interests include wireless communications and signal processing theory, including chaotic spread spectrum and LPI communications, physical layer security, spatial modulation,  deep learning and its applications in communication sensing, etc.
\end{IEEEbiography}

\begin{IEEEbiography}[{\includegraphics[width=1in,height=1.25in,clip,keepaspectratio]{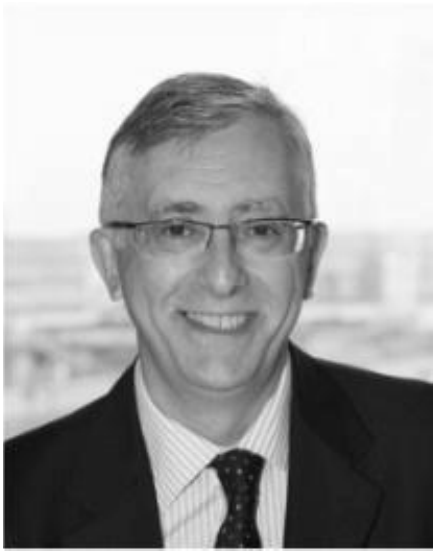}}]{Vincent Fusco}
(S'82-M'82-SM'96-F'04) received the bachelor's degree (Hons.) in electrical and electronic engineering, the Ph.D. degree in microwave electronics, and the D.Sc. degree from Queen's University Belfast (QUB), Belfast, U.K., in 1979, 1982, and 2000, respectively. 

He is currently a Chief Technology Officer with the Institute of Electronics, Communications and Information Technology (ECIT), QUB. He has authored more than 450 scientific papers in major journals and in referred international conferences and two textbooks. He holds patents related to self-tracking antennas and has contributed invited papers and book chapters. His current research interests include advanced front-end architectures with enhanced functionality, active antenna, and front-end MMIC techniques. 

Dr. Fusco is a Fellow of the Institution of Engineering and Technology, the Royal Academy of Engineers, and the Royal Irish Academy. He was a recipient of the IET Senior Achievement Award and the Mountbatten Medal, in 2012. He serves on the Technical Program Committee of various international conferences, including the European Microwave Conference.
\end{IEEEbiography}

\begin{IEEEbiography}[{\includegraphics[width=1in,height=1.25in,clip,keepaspectratio]{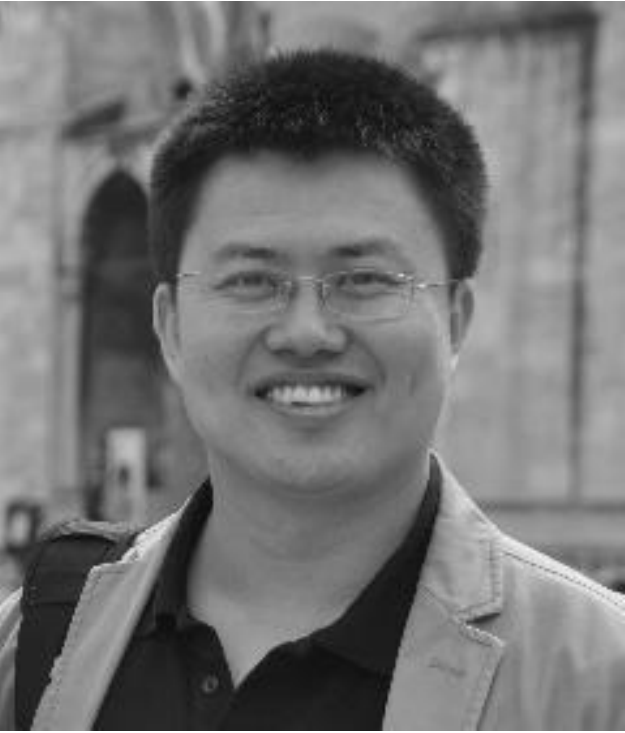}}]{Fanggang Wang}
(S'10-M'11-SM'16) received the B.Eng. and Ph.D. degrees from the School of Information and Communication Engineering, Beijing University of Posts and Telecommunications, Beijing, China, in 2005 and 2010, respectively. He was a Post-Doctoral Fellow with the Institute of Network Coding, The Chinese University of Hong Kong, Hong Kong, from 2010 to 2012. He was a Visiting Scholar with the Massachusetts Institute of Technology from 2015 to 2016, and with the Singapore University of Technology and Design in 2014. 

He is currently a Professor with the State Key Laboratory of Rail Traffic Control and Safety, School of Electronic and Information Engineering, Beijing Jiaotong University. His research interests are in wireless communications, signal processing, and information theory. He served as an Editor for the IEEE Communications Letters and a technical program committee member for several conferences.
\end{IEEEbiography}

\begin{IEEEbiography}[{\includegraphics[width=1in,height=1.25in,clip,keepaspectratio]{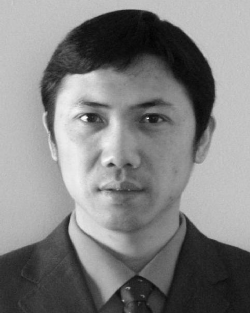}}]{Jiang Zhu}
received the B.S., M.S., and Ph.D. degrees in electrical engineering from the National University of Defense Technology (NUDT), Changsha, China, in 1994, 1997, and 2000, respectively. From 2000 to 2004, he was a lecturer in communication engineering at the NUDT. He was a visiting scholar at the University of Calgary, AB, Canada from April 2004 to July 2005. From 2005 to 2008, he was an associate professor in communication engineering at the NUDT. 

Since 2008, he has been with the NUDT as a full professor in the School of Electronic Science and Engineering. His current research interests include wireless high speed communication technology, satellite communication, physical layer security and wireless sensor network.
\end{IEEEbiography}

\begin{IEEEbiography}[{\includegraphics[width=1in,height=1.25in,clip,keepaspectratio]{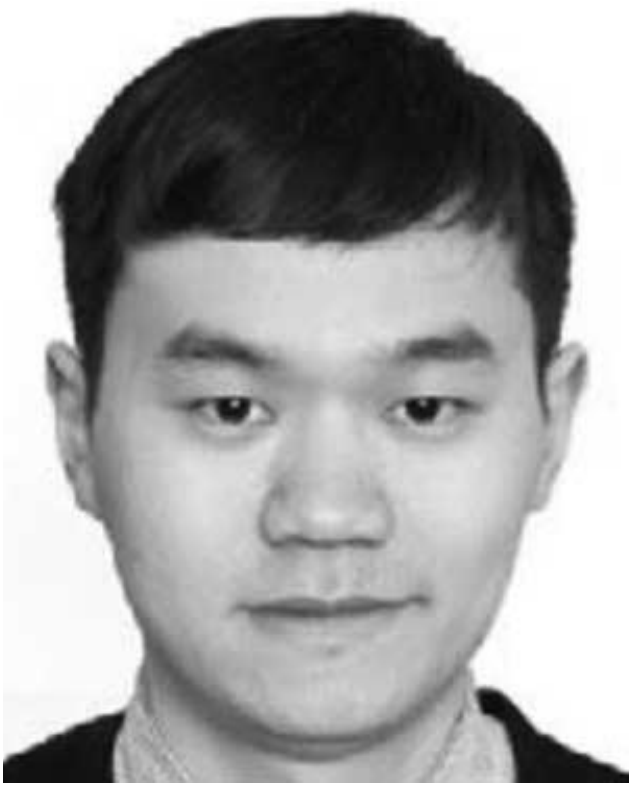}}]{Chao Gu}
received the B.S. and M.S. degrees in electronic engineering from Xidian University, Xi'an, China, in 2009 and 2012, respectively, and the Ph.D. degree in electronic engineering at University of Kent, Canterbury, U.K., in 2017. He is currently a senior engineer at the ECIT Institute, Queen's University Belfast, Belfast, U.K. His research interests include smart antennas, reconfigurable antennas, and frequency selective surfaces.
\end{IEEEbiography}

\end{document}